\documentclass[12pt, epsfig]{article}
\usepackage{amsmath,amsfonts,amssymb,amsthm,amstext,amscd,eucal,graphicx , xcolor}
\usepackage{slashbox}
\usepackage[all]{xy}
\usepackage{dsfont}
\usepackage{hyperref}
\usepackage{amsmath}
\usepackage{slashed}
\usepackage{caption}

\makeatletter \@addtoreset{equation}{section}

\makeatletter\renewcommand\section{\@startsection {section}{1}{\z@}%
                                   {-3.5ex \@plus -1ex \@minus -.2ex}%nn
                                   {2.3ex \@plus.2ex}%
                                   {\normalfont\large\bfseries}}
\renewcommand\subsection{\@startsection{subsection}{2}{\z@}%
                                     {-3.25ex\@plus -1ex \@minus -.2ex}%
                                     {1.5ex \@plus .2ex}%
                                     {\normalfont\bfseries}}

\newcommand{\be}{\begin{equation}}
\newcommand{\ee}{\end{equation}}
\newcommand{\bea}{\begin{eqnarray}}
\newcommand{\eea}{\end{eqnarray}}
\newcommand{\bse}{\begin{subequations}}
\newcommand{\ese}{\end{subequations}}
\newcommand{\beqa}{\begin{eqnarray}}
\newcommand{\eeqa}{\end{eqnarray}}
\newcommand{\beqar}{\begin{eqnarray*}}
\newcommand{\eeqar}{\end{eqnarray*}}
\newcommand{\bi}{\begin{itemize}}
\newcommand{\ei}{\end{itemize}}
\newcommand{\bn}{\begin{enumerate}}
\newcommand{\en}{\end{enumerate}}
\newcommand{\ba}{\begin{array}}
\newcommand{\ea}{\end{array}}
\newcommand{\bc}{\begin{center}}
\newcommand{\ec}{\end{center}}

\newcommand{\xdownarrow}[1]{%
	{\left\downarrow\vbox to #1{}\right.\kern-\nulldelimiterspace}}

\definecolor{darkgreen}{rgb}{0,0.3,0}
\definecolor{darkblue}{rgb}{0,0,0.3}
\definecolor{darkred}{rgb}{0.7,0,0}

\newcommand{\old}[1]{}%{\sout{#1}}

\begin{document}

\begin{titlepage}

\begin{flushright}\vspace{-3cm}
{
%{\tt arXiv:yymm.nnnn} \\
%IPM/P-2016/nnn  \\
\today }\end{flushright}
\vspace{-.5cm}

\begin{center}

\large{\bf{Ultra-Spinning Gauged Supergravity Black Holes \\ and their Kerr/CFT Correspondence}}

\bigskip\bigskip

\large{\bf{S.M. Noorbakhsh\footnote{e-mail:
m.noorbakhsh@semnan.ac.ir}$^{ }$
and  M. Ghominejad\footnote{e-mail: mghominejad@semnan.ac.ir }$^{}$  }}
\\

\vspace{5mm}
\normalsize
\bigskip\medskip
{\it Department of Physics, Semnan University, P.O. Box 35195-363, Semnan, Iran}
\smallskip
\\
%\date{\today}

\end{center}
\setcounter{footnote}{0}

\begin{abstract}

\noindent
Recently a new class of asymptotically AdS ultra-spinning black holes has been constructed with a noncompact horizon of finite area \cite{HennigarKubiznakMann:2014}, in which the asymptotic rotation is effectively boosted to the speed of light. We employ this technique for four-dimensional $U(1)^4$ and five-dimensional $U(1)^3$ gauged supergravity black holes. The obtained new exact black hole solutions for both cases possess a noncompact horizon; their topologies are a sphere with two punctures. We then demonstrate that the ultra-spinning limit commutes with the extremality condition as well as the near horizon limit for both black holes. We also show that the near horizon extremal geometries of the resulting ultra-spinning gauged supergravity black holes lead to the well-known result which contains an AdS$_2$ throat. We then obtain the $[(d-1)/2]$ central charges of the dual CFTs. By assuming the Cardy formula, we show that despite the noncompactness of the horizon, microscopic entropy of the dual CFT is precisely equivalent to the Bekenstein-Hawking entropy. 
\end{abstract}

\end{titlepage}
\renewcommand{\baselinestretch}{1.1}  %Line spacing
%%%%%%%%%%%%%%%%%%%%%%%%%%%%%%%%%%%%%%%%%%%%%%%%%%%%%%%%%%%%%%%%%%%%%%%%%%%%%%%%%%%%%%%%%%%
\tableofcontents
\section{Introduction} 
 The first solution of Einstein field equations which is their only $4$D spherically symmetric vacuum solution, describing the first black hole was given by the Schwarzschild metric. Moreover many other black hole solutions have been found. Black holes possess a special null surface called the event horizon that is generated by a null Killing vector field, where no object behind it can escape to infinity. Bekenstein discovered that to prevent the violation of the second law of thermodynamics in the presence of a black hole, it must be viewed as a thermodynamic object with entropy. Hawking showed that the black hole acts as a black body with finite temperature that can radiate away its mass. The  thermodynamic behavior of black holes points to the existence of an underlying black hole microstate structure. One of the main curiosities is How these microstates can explain the macroscopic Bekenstein-Hawking entropy from a statistical viewpoints. A complete answer has not been given yet, but in string theory, for a large class of extremal supersymmetric black holes, this question has been answered \cite{StromingerVafa}: Bekenstein-Hawking entropy can indeed be reproduced from a statistical entropy as the logarithm of the degeneracy of BPS states indeed \cite{Sen}. Recently the horizon fluffs proposal was demonstrated in \cite{AfsharGrumillerSheikhJabbari} to identify the microstates of three-dimensional Ba\~nados--Teitelboim--Zanelli (BTZ) 
 black holes as states that are marked by the conserved charges related to the nontrivial diffeomorphisms on the near horizon region. Also this proposal evaluated in \cite{SheikhJabbariYavartanoo} for the general AdS$_3$ black holes in the class of Ba\~nados geometries.
 
 The Kerr/CFT correspondence \cite{GuicaHartmanStrominger-KerrCFT} provides a rich setup, supporting the idea that whatever the states of quantum gravity are in the near horizon region of an extremal Kerr-AdS black hole, they are holographically dual to quantum states of a two-dimensional chiral (left-moving part) CFT. Since then many further examples \cite{HartmanStrominger:2009CFTDual,ChowCvetic-CFT:2008} have been investigated for a large class of black hole solutions. In all cases the statistical microscopic entropy of the dual CFT using the Cardy formula \cite{Cardy} precisely agrees with the Bekenstein-Hawking entropy of the black holes. 
 
 There is a classification of black hole solutions based on their horizon topology. In particular, the famous Hawking theorem for stationary, asymptotically flat vacuum Einstein black hole solutions in four dimensions asserts that their horizon topologies necessarily are $S^2$. By relaxing some of the Hawking's theorem assumptions, one may comes up with the some classes of black hole solutions with different horizon topology, such as $S^3$ and $S^2 \times S^1$ (black ring solutions) horizon topologies in five-dimension. Also in asymptotically AdS spacetimes, the horizon of a black hole may have a compact Riemann surface of any genus $g$ instead of spherical horizons \cite{Non-sphericalHorizon}, as well as black ring solutions with horizon topology $S^1 \times S^{d-3}$. In the case of adding a rotation to them, the horizon would then be noncompact and the obtaining geometries would describe a rotating black membranes with horizon topology $\mathbb{H} \times S ^{d-4}$. It was shown in four-dimension in \cite{GneccchiHristovKlemm:2014} and elaborated upon in \cite{Klemm:2014} that in Einstein-Maxwell-Lambda theory or more generally, in the presence of a scalar potential for $\mathcal{N}=2$ gauged supergravity, one can obtain black hole solutions with non-compact horizon topology with finite area, which can be topologically viewed as spheres with two punctures. Other examples of this kind of topology were constructed in \cite{HennigarKubiznakMann:2014,HennigarKubiznakMann:2015} by performing an ultra-spinning (super-entropic) limit to Kerr-AdS and $d$-dimensional multi-spinning Myers-Perry black holes. 
 
 The first effort to study ultra-spinning black holes was made by Emparan and Myers \cite{EmparanMyers:2003} for exploring the stability of Myers-Perry black holes at any arbitrarily large angular momentum with keeping mass fixed, in which the rotation parameter $a \rightarrow \infty$. One result is that area decreases as angular momentum increases, and in higher dimensional $d$, area shrinks to zero. By capturing the null geodesics in the plane of rotation one can realize that the horizon of this kind of ultra-spinning black hole is highly spread out in the plane of rotation while it shrinks in the perpendicular direction. Also it was argued that a Gregory-Laflamme-type instability can be seen for these solutions. The analogue limit can be used in the case of rotating AdS black holes in $d \ge 6$ by taking the rotation parameter approaches the AdS radius $\ell$, which gives the geometry of a black membrane by keeping the mass of the black hole fixed \cite{CaldarelliEmparan:2008}.  In \cite{ArmasObers} another technique was proposed by which one can also perform $a \rightarrow \infty$ while restricting the ratio $a/\ell$ to remain fixed. Furthermore, Caldarelli et al. constructed a new type of solution with horizon topology $\mathbb{H}^2 \times S^{d-4}$ by keeping the horizon radius $r_+$ fixed while zooming in to the pole and taking $a \rightarrow \ell$, which is called hyperboloid membrane limit \cite{CaldarelliEmparan:2008,CaldarelliLeigh:2011}. Recently a simple ultra-spinning (super-entropic) limit was introduced in \cite{HennigarKubiznakMann:2014,HennigarKubiznakMann:2015} for Kerr-Newman-AdS and $d$-dimensional multi-spinning Myers-Perry black holes. This technique begins with the Kerr-AdS black hole in an asymptotically rotating frame, and then boosts this rotation to the maximum value $a \rightarrow\ell$, while keeping the metric finite by introducing a change to corresponding azimuthal coordinate, and finally one can compactify this new coordinate to generate a new rapid black hole solution. The resulting black hole has a non-compact horizon but finite area, which is an interesting feature. Also it was shown in the context of an extended thermodynamic phase space where the cosmological constant can vary \cite{KastorRayTraschen:2009}that the entropy of some of these new solutions violates the reverse isoperimetric inequality, so such black holes are called "super-entropic" \cite{HennigarKubiznakMann:2014,HennigarKubiznakMann:2015}.\footnote[1]{It was shown in \cite{CveticGibbonsKubiznak-ISO:2012} that for a black hole of a given thermodynamic volume, the entropy inside a horizon is saturated for a (charged) Schwarzschild-AdS black hole. If the entropy of a black hole exceeds its expected maximal entropy, it will be denoted as super-entropic.} 
 
 The aim of this work is to find new rotating black hole solutions by taking the recent ultra-spinning (super-entropic) limit \cite{HennigarKubiznakMann:2014} on some existing black holes. The interesting geometry of resulting ultra-spinning black holes motivated us to further explore the applicability of this limit for some gauged supergravity black holes in four and five dimensions to generate a new type of black hole solutions. Since Supergravity solutions correspond to consistent string theory backgrounds, microscopic degeneracy of states can be investigated using the AdS/CFT correspondence, providing more motivation for their study in ultra-spinning limit. We will focus specifically on two gauged supergravity solutions in $\mathcal{N}=2$ and $\mathcal{N}=4$. After that the Kerr/CFT correspondence for obtaining gauged supergravity ultra-spinning black holes are investigated. It was shown in \cite{Mann-SEBH-CFT:2015} that despite the noncompactness of event horizons there exists this correspondence for the ultra-spinning Kerr-Newmann-AdS black hole as well as the ultra-spinning limit of minimal gauged supergravity black holes in five dimension.
 
 This paper is organized as follows. In Sec. II, we consider $U(1)^4$ gauged supergravity black hole in four dimensions and we discuss the geometry of the obtained ultra-spinning limit version. We then explore extremality conditions and the near horizon limit under the ultra-spinning limit for this black hole, indicating that both of them commute with the ultra-spinning limit. Next we provide a brief review of Kerr/CFT correspondence, followed by finding microscopic entropy via the Cardy formula, which is exactly equal to the black hole entropy. In Sec. III we have presented a similar analysis for multispinning $U(1)^3$ gauged supergravity black hole in five dimensions. Our conclusions with some remarks are given in Section IV.

\section{Four-Dimensional $U(1)^4$ Gauged Supergravity With Pairwise Equal Charge}\label{section-4d}
Here, we consider a four-dimensional rotating gauged supergravity black hole with pairwise-equal charged, which was constructed first in \cite{ChongCveticLuPope-4dgaugedBH} as a $U(1)\times U(1)$ Abelian subgroup of the $SO(4)$ gauged $\cal {N}$$=4$ supergravity. This solution can be consistently embedded in $\cal{N}$$=8$ gauged supergravity, in which two independent electromagnetic charges can be carried by fields in $U(1)$ subgroups of two $SU(2)$ sectors in $SO(4) \sim SU(2) \times SU(2)$.  Its relevant bosonic Lagrangian as a truncation of the $\mathcal{N}=4$ is given by
\begin{eqnarray}\label{Action4d}\nonumber
\mathcal{L}_4&=& R * \mathds{1} - \frac{1}{2} * d \varphi \wedge d \varphi - \frac{1}{2} e^{2 \varphi} * d \chi \wedge d \chi - \frac{1}{2} e^{- \varphi} * F_{_{(2)2}} \wedge F_{_{(2)2}} - \frac{1}{2} \chi F_{_{(2)2}} \wedge F_{_{_{(2)2}}} \\ \nonumber
&-& \frac{1}{2(1+\chi^2 e ^{2\varphi})} \big(e^{\varphi} * F_{_{(2)1}}\wedge F_{_{(2)1}} - e^ {2 \varphi} \chi \, F_{_{(2)1}} \wedge F_{_{(2)1}} \big) \\ 
&-&g^2\big(4+ 2 \cosh \varphi + e^{\varphi} \chi^2\big)* \mathds{1},
\end{eqnarray}
where $\varphi$ is the dilaton and $\chi$ is axion.\footnote[1]{ In \cite{ChongCveticLuPope-4dgaugedBH} a formalism has been proposed in ungauged $\mathcal{N} = 2$ supergravity coupled to three vector multiplets for generating the $4d$ solutions with four independent charges. The gauged pairwise-equal charges solution comes up by subtraction of the scalar potential from the ungauged bosonic Lagrangian with $\varphi_2=\varphi_3=\chi_2=\chi_3=0$.}
Indeed this special solution is a $U(1)^2$ subset of $U(1)^4$ which arises by setting two electric charges equal $(\delta_2=\delta_4)$, and taking the two magnetic charges equal $(\delta_1=\delta_2)$ as well. For more details we refer the reader to \cite{ChongCveticLuPope-4dgaugedBH}.

Four field strengths can be written in terms of potentials as
\begin{eqnarray}
F_{_{(2)1}} &=& d A_{_{(1)1}}, \quad \quad F_{_{(2)2}} = d A_{_{(1)2}},
\end{eqnarray}
and $g$ denotes the gauge-coupling constant, which is related to the AdS radius $\ell$ by $g = \ell^{-1}$. 
Its non-extremal black hole solution in asymptotic rotating frame (ARF) is given by \cite{ChongCveticLuPope-4dgaugedBH}
\begin{eqnarray}\label{Metric4d)}\nonumber
ds^2&=&-\frac{\Delta_r}{W}\big(dt - \frac{a \sin^2\theta}{\Xi} d \phi \big)^2+ W\big(\frac{dr^2}{\Delta_r}+\frac{d\theta^2}{\Delta_\theta}\big)+\frac{\Delta_\theta \sin^2\theta}{W}\big(a dt-\frac{r_1r_2+a^2}{\Xi}d\phi\big)^2,\\
\end{eqnarray}
where
\begin{eqnarray}\label{Metric4d-2)}\nonumber
\Delta_r&=&r^2+a^2 -2m r+\frac{1}{\ell^2}r_1r_2(r_1r_2+a^2),\\
\Delta_{\theta}&=&1-\frac{a^2}{\ell^2}\cos^2\theta, \quad \quad  W=r_1r_2 + a^2 \cos^2\theta,\\\nonumber
r_i&=&r+2ms_i^2=r+q_i, \quad \quad \Xi=1-\frac{a^2}{\ell^2},\\ \nonumber
s_i&=&\sinh \delta_i, \quad \quad c_i=\cosh \delta_i.
\end{eqnarray}
Also the axion, dilaton and gauge potentials read
\begin{eqnarray}
e^{\varphi}&=&\frac{r_1^2+a^2\cos^2\theta}{W}, \quad \quad \quad \quad \quad \quad \qquad  \quad \quad \quad \chi=\frac{a(r_2-r_1)\cos^2\theta}{r_1^2+a^2\cos^2\theta},\\ \nonumber
A_{_{(1)1}}&=&\frac{2\sqrt{2}m (dt-a \sin^2\theta \Xi^{-1} d \phi)}{W} s_1 c_1 r_2, \quad  \quad \quad A_{_{(1)2}}=\frac{2\sqrt{2}m (dt-a \sin^2\theta \Xi^{-1} d \phi)}{W} s_2 c_2 r_1.\nonumber
\end{eqnarray}
The coordinate change $\tilde{\phi}=\phi+a g^2 t$ yields an asymptotically static frame (ASF). The Hawking temperature, entropy, angular velocity and electrostatic potentials on the horizon (in the asymptotically rotating frame) are
\begin{eqnarray}\label{Thermo4d}\nonumber
T_H&=&\frac{r_+^2-a^2+a^2/\ell^2(r_+^2-q_1q_2)+(r_++q_1)(r_++q_2)(3r_+^2+q_1r_+ + q_2 r_+ - q_1q_2)/\ell^2}{4 \pi r_+ [(r_++q_1)(r_++q_2)+a^2]},\\ \nonumber
S&=&\frac{\pi[(r_++q_1)(r_++q_2)+a^2]}{\Xi}, \quad \quad \quad \quad \quad \quad  \Omega=\frac{\Xi\, a}{(r_++q_1)(r_++q_2)+a^2}, \\
\Phi_1&=&\Phi_2=\frac{2ms_1 c_1(r_++q_2)}{(r_++q_1)(r_++q_2)+a^2}, \quad\qquad\qquad  \Phi_3=\Phi_4=\frac{2ms_2 c_2(r_++q_1)}{(r_++q_1)(r_++q_2)+a^2},
\end{eqnarray}
where $r_+$ is the largest root of $\Delta_r=0$ as the outer horizon. 

The charge quantities including mass, angular momentum and pairwise electric potentials were constructed in \cite{CvetiGibbons:2005} and are
\begin{eqnarray}\label{Charges4d}\nonumber
E&=&\frac{m}{\Xi^2}(1+s_1^2+s_2^2)=\frac{2m+q_1+q_2}{2 \, \Xi^2},  \\ \nonumber
J&=&\frac{m a}{\Xi^2}(1+s_1^2+s_2^2)=\frac{a(2m+q_1+q_2)}{2 \, \Xi^2},\\ 
Q_1&=&Q_2=\frac{ms_1 c_1}{2\Xi}, \quad\quad\quad \quad Q_3=Q_4=\frac{m s_2 c_2}{2 \Xi}.
\end{eqnarray}
In the next subsection we use the ultra-spinning (super-entropic) limit upon the metric (\ref{Metric4d)}) in order to obtain a new charged-AdS black hole solution in gauged supergravity.

This novel ultra-spinning (super-entropic) limit can be interpreted as a simple method to generate a new black hole solution in which the rotation parameter $a$ reaches its maximum amount, equal to the AdS radius $\ell$. This procedure limit consists of three steps. \newline
i) Transforming metric to an asymptotic rotating frame to avoid a singular metric in the ultra-spinning limit. We then need only define a new azimuthal coordinate $\varphi=\phi/{\Xi}$ afterward. \newline ii) This rotation has to be boosted effectively to the speed of light, namely, by taking the $a \rightarrow \ell$ limit. \newline
iii) In final step, we compactify the new azimuthal direction $\varphi$ \cite{HennigarKubiznakMann:2014}. 

We note that employing an asymptotically rotating coordinate should be assumed as a crucial point in the ultra-spinning limit technique. The uniqueness of the choice of ARF to have a nonsingular black hole solution has been discussed in \cite{HennigarKubiznakMann:2015}. In fact, avoiding to start from an ARF leads us to a singular limit. 

%%%%%%%%%%%%%%%%%%%%%%%%%%%%%%%%%%%%%%%%%%%%%%%%%%%%%
 \subsection{Ultra-spinning limit}
 Since the metric (\ref{Metric4d)}) is already written in an asymptotically rotating frame, we therefore need only to introduce a new azimuthal coordinate $\varphi=\phi/\Xi$, followed by taking the limit $a \rightarrow \ell$. Hence we straightforwardly obtain the following solution
 \begin{equation}\label{UsMetric4d}
 ds^2=-\frac{\tilde{\Delta}_r}{\tilde{W}}\big(dt - \ell \sin^2\theta d \varphi \big)^2+ \tilde{W}\big(\frac{dr^2}{\tilde{ \Delta}_r}+\frac{d\theta^2}{\sin^2\theta}\big)+\frac{\sin^4\theta}{\tilde{W}}\big[\ell dt-(r_1r_2+\ell^2 )d\varphi\big]^2,
 \end{equation}
 where $\tilde{\Delta}_r$ and $\tilde{W}$ are given by (\ref{Metric4d-2)}) as $a \rightarrow l$. To exclude a conical singularity in $\varphi$ direction, one can identify it with period $2\pi/\Xi$. Since the new azimuthal coordinate $\varphi$ is non-compact, we now compactify it by requiring 
 \begin{equation}
 \varphi \sim \varphi + \mu,
 \end{equation}
 where the parameter $\mu$ is dimensionless. Note that there is still an axial Killing vector $\partial_\varphi$ in the new coordinate direction $\varphi$. Therefore, it is straightforward to show that the obtained metric (\ref{UsMetric4d}) appears as a new exact solution. Also we can easily find the dilaton, axion and gauge fields in this limit as
 \begin{eqnarray}\label{fieldsUS4d}
 e^{\tilde{\varphi}}&=&\frac{r_1^2+\ell^2\cos^2\theta}{\tilde{W}}, \quad \quad \quad \quad \quad \quad \quad \quad \quad \quad \tilde{\chi}=\frac{\ell(r_2-r_1)\cos\theta}{r_1^2+\ell^2\cos^2\theta},\\ \nonumber
 \tilde{A}_{_{(1)1}}&=&\frac{2\sqrt{2}m s_1 c_1 r_2 (dt-\ell \sin^2\theta  d \varphi)}{\tilde{W}}, \quad \quad \quad  \tilde{A}_{_{(1)2}}=\frac{2\sqrt{2}m s_2 c_2 r_1 (dt-\ell \sin^2\theta  d \varphi)}{\tilde{W}}. 
 \end{eqnarray}
 
 %%%%%%%%%%%%%%%%%%%%%%%%%%%%%%%%%%%%%%%%%%%%%%%%%%%%%%%%%%%%%%%%%%%%%%%%%%%%%%%%%%%%%%%%%%%%%%%%%%%%%%%%%%%%%%%%%%
 \textbf{Horizon geometry: }
 To ensure that the new solution (\ref{UsMetric4d}) is describing a black hole, we examine the largest root of $\tilde{\Delta}_r$ in (\ref{UsMetric4d}), which is supposed to demonstrate the location of the horizon as $r_+$. It is required now to check $\tilde{\Delta}^{\prime}_r \geq 0 $. We therefore find following mass bound 
 \begin{eqnarray}\label{mass4d}
 m &\geq& m_0\equiv \frac{\sqrt{3}}{18 \ell^2}\big[(q_1-q_2)^2-4\ell^2\big]^{\frac{3}{2}}-\frac{q_1+q_2}{2}, \quad \quad |q_2-q_1| \geq 2 \ell.
 \end{eqnarray}
 For $m > m_0$ a horizon exists, while for $m < m_0$ there exists a naked singularity. Also to ensure that our obtained geometry would be free of any closed timelike curves (CTC) we examine $g_{\varphi\varphi} \ge 0$ 
 \begin{eqnarray}
  g_{\varphi\varphi}=\frac{\ell^2\sin^4\theta\big[(2 m+ r_+ (q_1+q_2)+q_1q_2\big] }{ (r_+ +q_1)(r_++q_2) + \ell^2 \cos^2\theta}.
 \end{eqnarray}
 Apparently $g_{\varphi\varphi}$ is positive in the entire spacetime.
 
 Now, let us take a deeper look at the geometry of the horizon. The induced metric on a constant $(t,r)$ surface yields,
 \begin{eqnarray}\label{ds2h-1}
 ds^2_h=\frac{\tilde{W}_+}{\sin^2\theta}d\theta^2+\sin^4\theta\frac{\big((r_+ +q_1)(r_++q_2)+\ell^2\big)^2}{\tilde{W}_+}d\varphi^2.
 \end{eqnarray}
 where $\tilde{W}_+=\tilde{W}|_{r_+}$. However, this geometry seems to be singular in $\theta=0$ and $\pi$, so we will show that the symmetry axis $\theta=0,\pi$ is actually not part of the spacetime. For a precise study about these poles, one can probe the metric in the small $\theta=0$ limit by introducing the change of variables similar to \cite{HennigarKubiznakMann:2015}, 
 \begin{equation} \label{definek}
 k=\ell(1-\cos\theta),
 \end{equation}s
 Horizon metric (\ref{ds2h-1}) for small $k$ becomes 
 \begin{eqnarray}\label{ds2h-2}
 ds^2_h=(r_+ +q_1)(r_++q_2)\bigg[\frac{dk^2}{4k^2}+\frac{4k^2}{\ell^2}d\varphi^2\bigg],
 \end{eqnarray}
 which shows clearly a metric of constant negative curvature on a quotient of the hyperbolic space $\mathbb{H}$$^2$. Not that, due to the symmetry, the $\theta=\pi$ limit gets the same result. Thus, there is no true singularity at these two points, but some sort of boundaries. Therefore topologically, the event horizon is a sphere with two punctures, which implies that our obtained black hole enjoys a finite area but non-compact horizon.
 
 In order to visualize the geometry of the horizon (\ref{ds2h-1}) we can embed it in Euclidean $3$-space as a surface of revolution \cite{GneccchiHristovKlemm:2014}. We then identify the induced metric on the horizon (\ref{ds2h-1}) by using a flat metric in cylindrical coordinates as
 \begin{equation} \label{CylindricalCo1}
 ds_3^2=dz^2+dR^2+R^2d \Phi^2,
 \end{equation}
 and consider $z = z(\theta)$, $R = R(\theta)$. Setting $\Phi=\frac{2\pi}{\mu} \varphi$, one gets
 \begin{equation}\label{CylindricalCo2}
 R^2(\theta)=\bigg(\frac{\mu}{2\pi}\bigg)\frac{[(r_++q_1)(r_++q_2)+\ell^2]^2}{(r_++q_1)(r_++q_2)+\ell^2 \cos^2\theta}\sin^4\theta,
 \end{equation}
 \begin{equation}\label{CylindricalCo3}
 \bigg(\frac{dz(\theta)}{d\theta}\bigg)^2 + \bigg(\frac{dR(\theta)}{d\theta}\bigg)^2=\frac{(r_++q_1)(r_++q_2)+\ell^2 \cos^2\theta}{\sin^2\theta},
 \end{equation}
 which is a differential equation for $dz/d\theta$. We integrated (\ref{CylindricalCo3}) numerically for the values $\ell=1$, $\mu=2\pi$, and $r_+=0.8$, by choosing different amounts of $q_1$ and $q_2$. The resulting surfaces of revolution are shown in Fig. 1. 
 \begin{figure}[htp]
 	\centering
 	\begin{tabular}{ccc}
 		\includegraphics[width=0.3\textwidth,height=0.3\textheight]{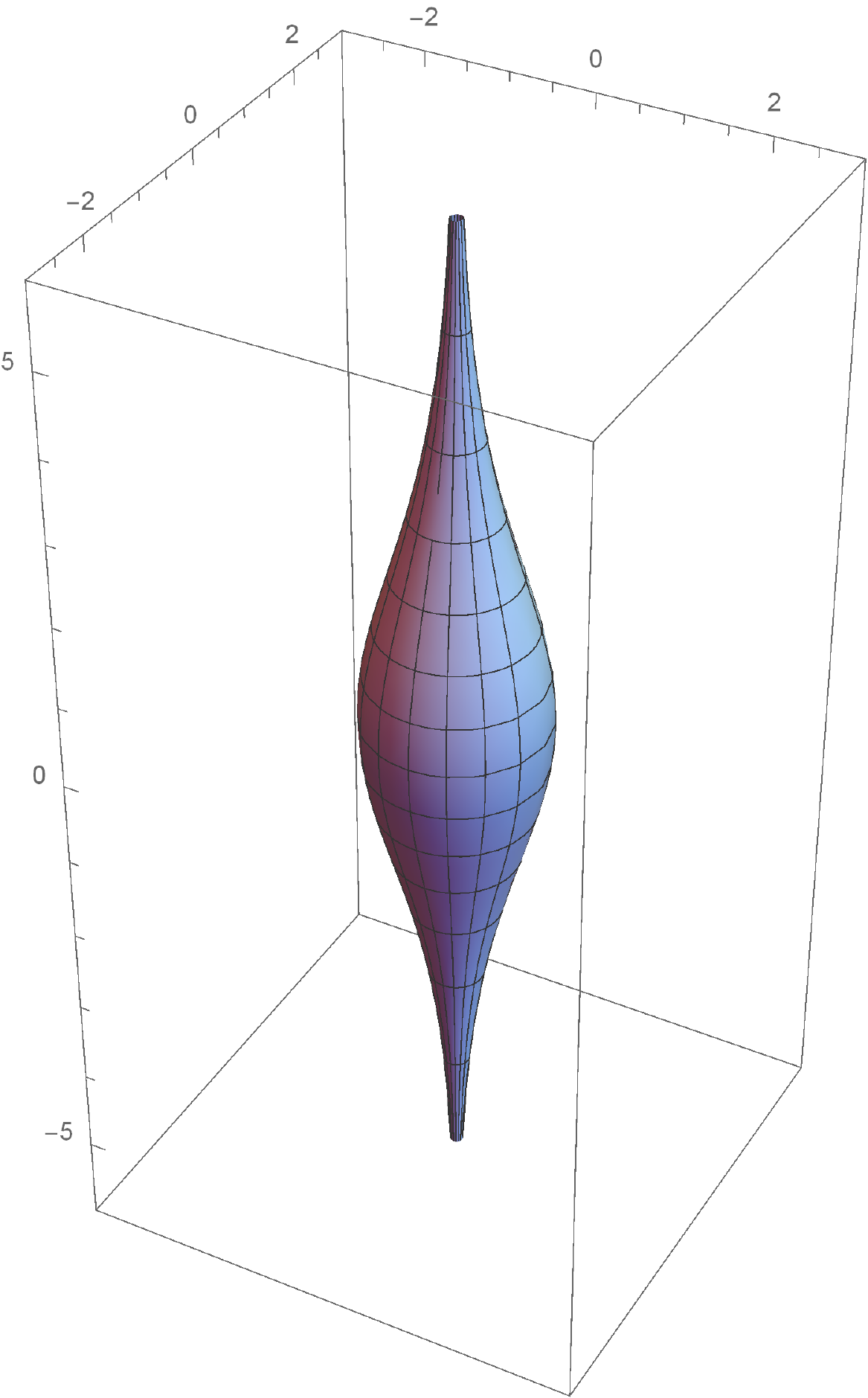} &
 		\includegraphics[width=0.3\textwidth,height=0.3\textheight]{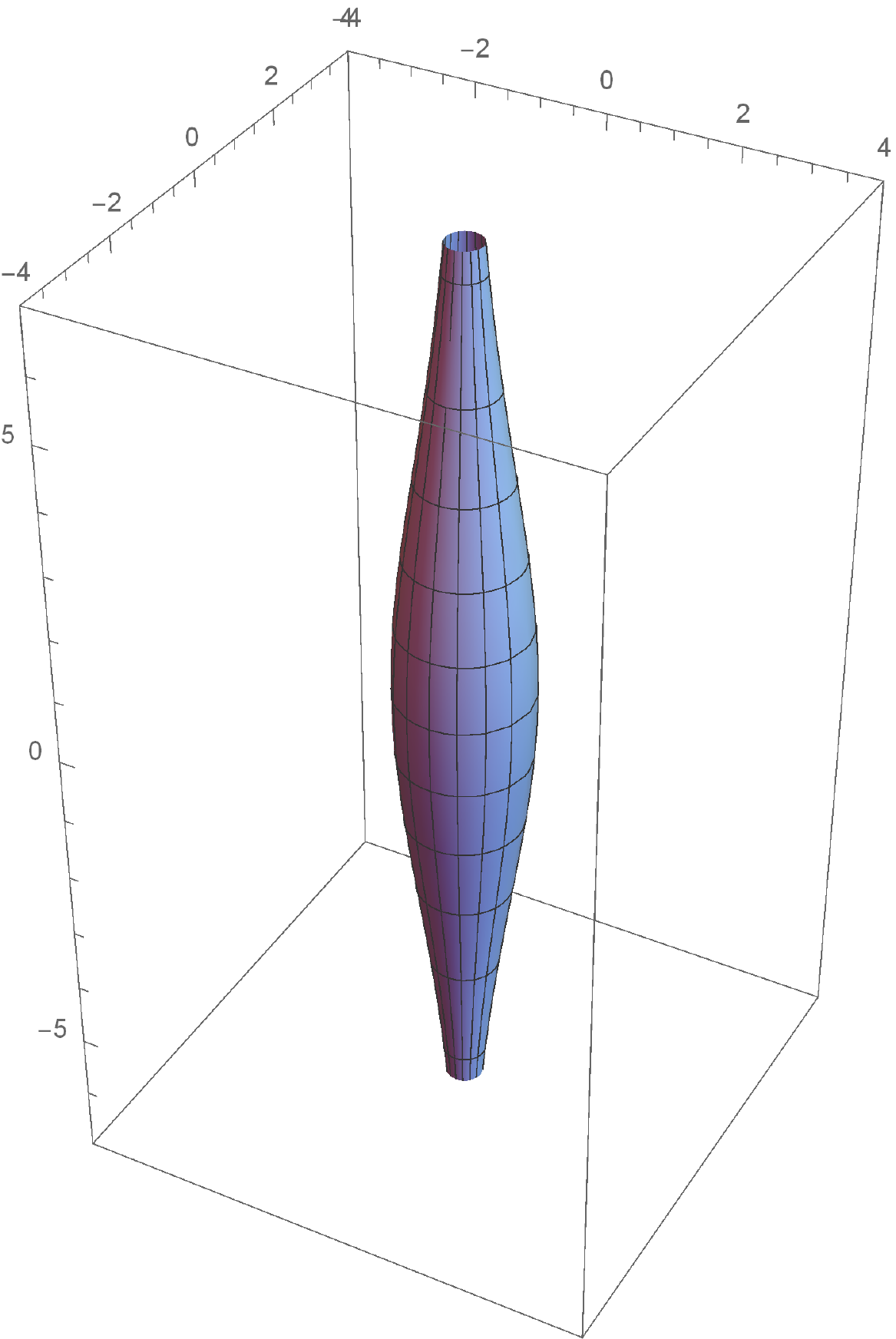}&
 		\includegraphics[width=0.3\textwidth,height=0.3\textheight]{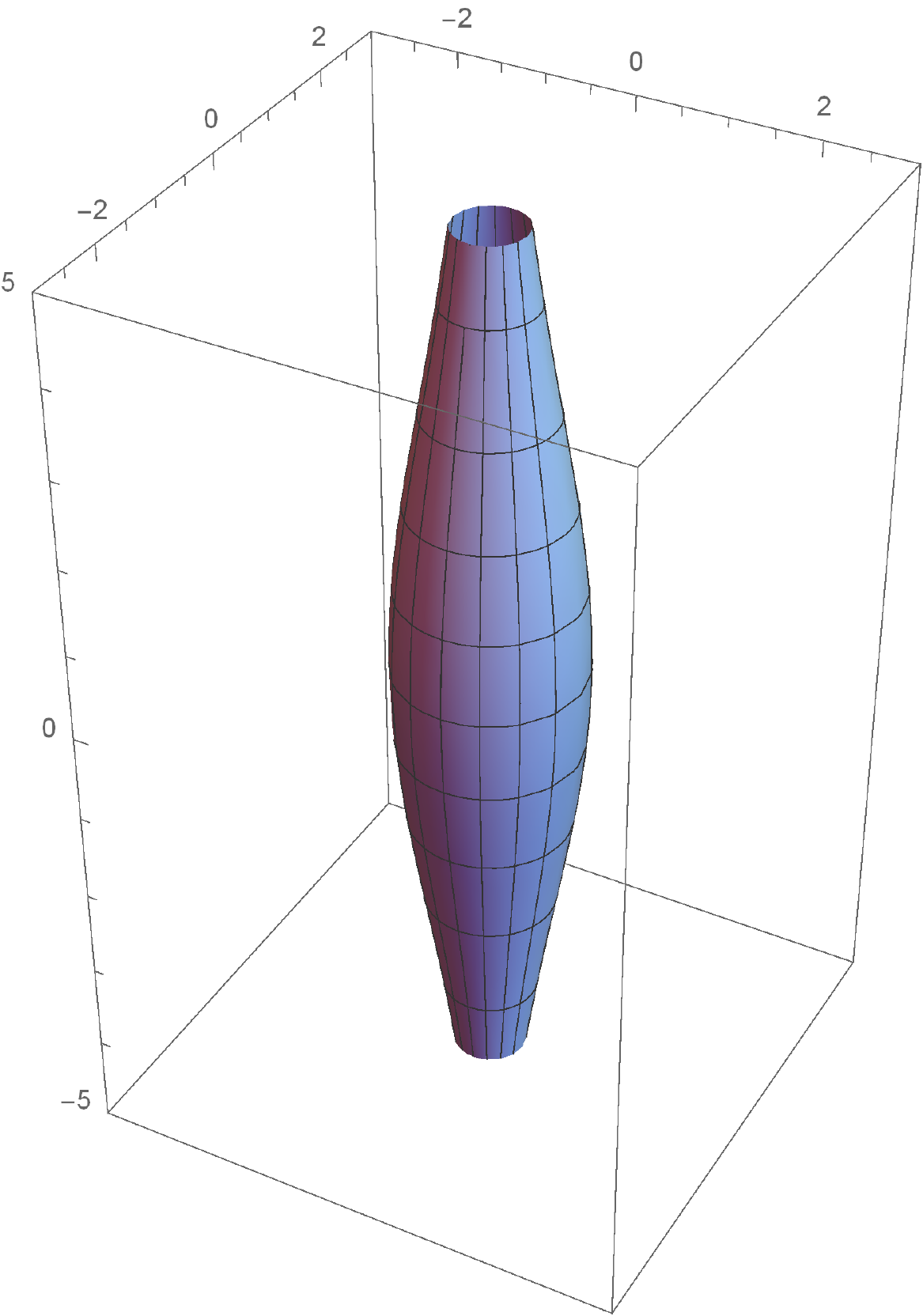}\\
 	\end{tabular}
 	\captionsetup{labelformat=empty}
 	\caption{{\bf Fig. 1. \bf Horizon embeddings in 4d}. Plots show two-dimensional nonacompact horizons embedded in $\mathbb{R}$$^3$ as a surface of revolution, displaying the topology of a sphere with two punctures. We have set $\mu = 2\pi$, $r_+ = 0.8$,  and $\ell=1$ for all diagrams, with  $q_1=10$, $q_2=0$ (left); $q_1=10$, $q_2=5 $ (middle); and $q_1=3$, \, $q_2=6$ (right).  }\label{HorizonEmbedding4d}
 \end{figure}\label{Fig1-4dH}

 %%%%%%%%%%%%%%%%%%%%%%%%%%%%%%%%%%%%Conformal Boundary%%%%%%%%%%%%%%%%%%%%%%%%%%%%%%%%%%%%%%%%%
 \textbf{Conformal Boundary :}
 Let us here, find the conformal boundary of our obtained new black hole solution (\ref{UsMetric4d}) (the conformal factor is $\ell^2/r^2$)
 \begin{equation}\label{ConformalbdryUS4d}
 ds^2_{bdry}=-dt^2+\ell\sin^2\theta dt d\varphi +\frac{1}{\sin^4\theta}d\theta^2.
 \end{equation}
 It appears that the new coordinate $\varphi$ would be a null one on the conformal boundary. Now we study this metric near the pole $\theta=0$ using (\ref{definek}). For small $k$ we have then
 \begin{equation}\label{ConformalbdryUS4d-k}
 ds^2_{bdry}=-dt^2 +2 k dt d\varphi+\frac{1}{4 k^2}dk^2.
 \end{equation}
 This metric can be interpreted as an AdS$_3$ written in Hopf-like fibration over $\mathbb{H}^2$. It means again that the poles $\theta=0, \, \pi$ are indeed not parts of the spacetime, and they are being removed from the boundary.
 However, to answer what happens precisely at $\theta=0$ and  $\pi$, we should study the behavior of the geodesics in the entire spacetime. To do that, it should be shown that no outgoing null geodesics from inside the horizon would not be able to reach to the symmetry axis $\theta=0$, in a finite affine parameter, similar to the strategy taken in \cite{HennigarKubiznakMann:2015}. This study of course, is not in the scope of the current paper, so we postpone studying it to another further work.
 
As we recall, the AdS/CFT viewpoint hints at exploring a dual $3$-dimensional CFT which exists on the AdS$_3$ conformal boundary (\ref{ConformalbdryUS4d-k}). We note, depending on the choice of the AdS coordinate slicing, different manifolds may be achieved such that one can found a dual CFT that lives on the correspondence slicing. Each coordinate indeed covers the whole or a part of AdS boundary; for instance, in global coordinate the dual CFT lives on $R \times S^p$ geometry, while in the Poincar$\acute{e}$ patch the CFT exists on $R^{p,1}$ manifold. Therefore, in the slicing (\ref{ConformalbdryUS4d-k}) one may expect the dual CFT to reside on an AdS$_3$ geometry. These CFTs can be viewed as possible deformations around the conformal fixed point, which may be related to each other by the Wilsonian renormalization group flow equation. 

 %%%%%%%%%%%%%%%%%%%%%%%%%%%%%%%%%%%%%%%%% Thermodynamic quantity %%%%%%%%%%%%%%%%%%%%%%%%%%%%%%%%%%%
 \textbf{Thermodynamic quantities :}
 Although, the event horizon of the obtained ultra-spinning black hole (\ref{UsMetric4d}) is noncompact, it has a finite area and entropy as
 \begin{equation}\label{EntropyUS4d}
 S =\frac{\mu}{2} \big[ (r_+ +q_1)(r_+ + q_2)+\ell^2 \big].
 \end{equation}
 For solution (\ref{UsMetric4d}) we extract the Hawking temperature, electrostatic potentials and angular velocity on the horizon. Our results are 
 \begin{eqnarray}\label{thermoUS4d}\nonumber
 T_H&=& \frac{2r_+^2-\ell^2-q_1q_2+(r_+ +q_1)(r_+ +q_2)(3r_+^2+q_1r_+ + q_2r_+-q_1q_2)/\ell^2}{4\pi r_+
 	\big[(r_++q_1)(r_++q_2)+\ell^2\big]},\\ \nonumber
 \Phi_1&=& \Phi_2 = \frac{\sqrt{q_1(2m+q_1)}(r_++q_2)}{(r_++q_1)(r_++q_2)+\ell^2}, \quad \quad \quad  \Phi_3= \Phi_4 = \frac{\sqrt{q_2(2m+q_2)} (r_++q_1)}{(r_++q_1)(r_++q_2)+\ell^2}, \\
 \Omega&=&\frac{\ell}{(r_+ +q_1)(r_+ + q_2)+\ell^2}.
 \end{eqnarray}
 The electric potential is computed using $\Phi^I=K^\mu A_\mu^I$, where $K^\mu=\partial_t+\Omega_s \partial_\varphi$ is the null Killing vector generating the horizon. We then calculated the following conserved charges as
 \begin{eqnarray}\label{ChargesUS4d}
 E&=&\frac{2m+q_1+q_2}{2}, \quad\quad\quad \quad \quad\quad\quad  \quad \quad\quad \quad \quad  J=\frac{\ell(2m+q_1+q_2)}{2},\\  \nonumber
 Q_1&=&Q_2=\frac{\sqrt{q_1(2m+q_1)}}{4}, \quad\quad \quad\quad  \quad\quad\quad Q_3=Q_4=\frac{\sqrt{q_2(2m+q_2)}}{4}.
 \end{eqnarray}
 Angular momentum can be obtained by using the Komar integral based on the Killing vector $\partial_\varphi$, also the electric charges are calculated by using the Gaussian integrals. For computing mass we use the conformal approach of Ashtekar, Magnon and Das (AMD) \cite{Ashtekar}. In the AMD method the electric part of the Weyl tensor plays an essential role: its spatial conformal boundary integral gives us the mass. It was shown that for a large variety of gauged supergravity black holes this mass exactly agrees with the first law of thermodynamics calculation \cite{Chen:2006:MassAMD} .
 %%%%%%%%%%%%%%%%%%%%%%%%%%%%%% Extremality under ultra-spinning limit %%%%%%%%%%%%%%%%%%%%%%%%%%
 
\textbf{Extremality under ultra-spinning limit: } For an extremal black hole the inner and outer horizons coincide to a single horizon at $r_0$, with vanishing Hawking temperature and generically non-zero entropy.
 
Here, in order to answer whether extremality condition will be preserved under the ultra-spinning limit or not, we consider two different approaches: \newline
i) Finding the extremality condition of a given black hole and then applying the ultra-spinning limit.\newline
ii) Performing the ultra-spinning limit to a given geometry, and then finding the extremality condition of the obtained ultra-spinning black hole afterwards.

 Firstly, let us to start from situation (i), the extremality conditions of metric (\ref{Metric4d)}) can be found by using (\ref{Thermo4d}) and (\ref{Metric4d-2)}), imposing $T_H|_{r=r_0}=0$ and $\Delta_r=0$. Hence the degenerate horizon $r_0$ reads
 \begin{eqnarray}\label{ExCondition4d-1}
  r_0=\frac{1}{6}\bigg[\big(3\mathcal{B}+3\sqrt{\mathcal{A}^3+\mathcal{B}}\big)^{1/3}-\frac{\mathcal{A}}{\big(3\mathcal{B}+3\sqrt{\mathcal{A}^3+\mathcal{B}}\big)^{1/3}}-3(q_1+q_2)\bigg],
 \end{eqnarray}
where $\mathcal{A}=3^{2/3}\big[2(a^2+\ell^2)-(q_1-q_2)^2\big]$, and $\mathcal{B}=9\ell^2(2m+q_1+q_2)$. Also the following constraints between parameters of solution and $r_0$ are achieved
 \begin{eqnarray}\label{ExCondition4d-2}
  q_1&=&\frac{a^2 q_2 -4 r_0^2(r_0+q_2)+\sqrt{4(r_0^2-q_2^2)[-a^2\ell^2 +r_0^2(a^2+\ell^2+q_2^2+4q_2r_0+3r_0^2) ]}}{2(r_0^2-q_2^2)}.
  \\ \nonumber
 m&=&\frac{a^4 q_2 + 4 \ell^2 q_2(r_0-q_2)+2 (r_0 + q_2)^2\big[2r_0(r_0+q_2)^2+\chi)\big]-a^2\big[4\ell^2(q_2-r_0)+4r_0(r_0+q_2)^2+\chi\big]}{4 \ell^2 (r_0-q_2)^2},
 \end{eqnarray}
 where $\chi=\sqrt{a^4q_2^2+4(r_0+q_2)\big(r_0^2 (q_2 + r_0) ((q_2 + r_0)^2 - a^2) + 4 l^2 (q_2 - r_0) (r_0^2 - a^2 ) \big)}$.

 Now, upon taking the limit $a \rightarrow \ell$, one easily gets
 \begin{eqnarray}\label{ExConditionUS4d-1}
 q_1&=&\frac{\ell^2 q_2 - 4 q_2 r^2 - 4 r^3+\sqrt{ 4 r_0^2 (q_2 + r_0)^4 + \ell^4 (-3 q_2^2 + 4 r_0^2)-8 \ell^2 r_0^3 (q_2 + r_0)}}{2(r_0^2-q_2^2)}, \\ \nonumber
 m&=&\frac{\ell^4 (-3 q_2 + 4 r_0) + 2 (q_2 + r_0)^2\big[2r_0(q_2+r_0)^2+\chi_s\big]-\ell^2\big[4r_0^2(3q_2+r_0)+\chi_s\big]}{4 \ell^2 (r_0-q_2)^2},
 \end{eqnarray}
 where $\chi_s=\sqrt{4r_0^2(r_0+q_2)^4+\ell^4(4r_0^2-3q_2^2)-8\ell^2r_0^3(r_0+q_2)}$. The horizon also obtained as
 \begin{eqnarray}\label{ExConditionUS4d-2}
 r_0=\frac{1}{6}\bigg[\big(3\mathcal{B}+3\sqrt{\tilde{\mathcal{A}}^3+\mathcal{B}}\big)^{1/3}-\frac{\tilde{\mathcal{A}}}{\big(3\mathcal{B}+3\sqrt{\tilde{\mathcal{A}}^3+\mathcal{B}}\big)^{1/3}}-3(q_1+q_2)\bigg],
 \end{eqnarray}
 where $\tilde{\mathcal{A}}=3^{2/3}\big[2\ell^2-(q_1-q_2)^2\big]$.
 
Now, we examine approach (ii). In this way we only need to find the extremality condition of our obtained ultra-spinning black hole (\ref{UsMetric4d}) by using (\ref{thermoUS4d}). Our resulting conditions are exactly same as (\ref{ExConditionUS4d-1}) and (\ref{ExConditionUS4d-2}). It means that the extremality conditions commute with the ultra-spinning limit for a  four-dimensional $U(1)^4$ gauged supergravity (\ref{Metric4d)}) black hole.

 %%%%%%%%%%%%%%%%%%%%%%%%%%%%%%%%%%%%%%%%%%%%%%%%%%%%%%%%%%%%%%%%%%%%%%%%%%%%%%%%%%%%%%%%%%%%%%%%%%%%%  
 %%%%%%%%%%%%%%%%%%%%%%%%%%%%%%%  Near Horizon Geometry    %%%%%%%%%%%%%%%%%%%%%%%%%%%%%%%%%%%
 %%%%%%%%%%%%%%%%%%%%%%%%%%%%%%%%%%%%%%%%%%%%%%%%%%%%%%%%%%%%%%%%%%%%%%%%%%%%%%%%%%%%%%%%%%%%%%%%%%%%%
 \subsection{Near horizon geometry}
 Extremal black holes have several interesting characteristics. One of them is that the near horizon region of extremal black holes intuitively can be interpreted as an isolated geometry and isolated thermodynamical system. It has been discussed in \cite{BardeenHorowitz,KunduriLucietti} that focusing on the near horizon geometry of extremal black holes leads to a new class of solutions to the same theory of gravity, which their conserved charges are the same as the ones of the original black hole. Near horizon extremal geometries (NHEG) have no horizon and no singularity unlike black holes, with enhanced $SL(2,\mathbb{R}) \times U(1)^N$ symmetry, and their asymptotic behaviours are different. Analogues to black hole thermodynamics, the laws of NHEG dynamics have been derived in \cite{Sheikh-JabbariHajianSeraj:2013}, in which despite the absence of a event horizon in NHEG there is an entropy associated to them as a Noether charge. However in this case system cannot be excited while it keeps $SL(2,R)$ isometry \cite{Reall,JohnstoneSheikh-Jabbari}. It was provided a proof in \cite{{Astefanesei, Astefanese2}} that $AdS_2$ sector appears in the near horizon geometry of any regular stationary extremal black hole. 
   
 In this section we try to find the near horizon geometry of the obtained ultra-spinning black hole  (\ref{UsMetric4d}) in the extremal limit. Now we can write the near horizon expansion as
 \begin{eqnarray}\nonumber
 \tilde{\Delta}_r=X(r-r_0)^2+\mathcal{O}(r-r_0)^3,
 \end{eqnarray}
 where
 \begin{eqnarray}
 X=2+\frac{1}{\ell^2}(6r_0^2+6q_1r_0+6q_2r_0+q_1^2+q_2^2+4q_1q_2).
 \end{eqnarray}
 Now, in order to find the near horizon geometry in the extremal limit, we use following dimensionless coordinate changes $(\hat{t},\hat{r},\hat{\theta},\hat{\varphi})$ to  extract the NHEG as an exact solution.
 \begin{eqnarray}\label{NHG-Coordinate4d}
 r&=&r_0(1 +\lambda  \hat{r}),\quad \quad \quad \varphi=\hat{\varphi}+\Omega_H^0 \hat{t}, \quad \quad \quad
 t=\frac{\hat{t}}{2\pi T^{\prime}_H r_0 \lambda },\quad \quad \quad \hat{\theta}=\theta.
 \end{eqnarray}
 Here, the quantity $\Omega_H^0$ is the angular velocity on the horizon in extremal limit, which is a shift for the coordinate $\varphi$ to attain the comoving coordinate of the horizon. In these new coordinates the killing vector $\xi=\lambda/r_0 \partial_t$ is the horizon generator. Also the quantity $T ^{\prime \, 0}_H$ is defined as
 \begin{equation}
 T^{\prime \, 0}_H=\left.\frac{\partial T_H}{\partial \hat{r}_+ }\right\vert_{\hat{r}_+=r_0}.
 \end{equation}
 
 Now upon taking the limit $\lambda \rightarrow 0$, the near-horizon metric reads,
 \begin{eqnarray}\label{NH4d}
 ds^2&=&\frac{\tilde{W}_0}{X}\big(-\hat{r}^2 d\hat{t}^2 + \frac{d \hat{r}^2}{\hat{r}^2}\big) + \frac{\tilde{W}_0}{\sin^2\theta}d\hat{\theta}^2,\\ \nonumber
 &+& \frac{\sin^4 \theta}{\ell^2 \tilde{W}_0}\big[(r_0+q_1)(r_0+q_2)+\ell^2\big]^2 \left( d\hat{\varphi}  + k \hat{r} d\hat{t}  \right)^2,
 \end{eqnarray}
 where 
 \begin{equation}\label{k4d}
 k=\frac{\ell (2r_0+q_1+q_2)}{X\big[(r_0+q_1)(r_0+q_2)+\ell^2\big]},
 \end{equation}
 and $\tilde{W}_0=\tilde{W}|_{\hat{r}=r_0}$. This metric can be viewed as the direct product of AdS$_2 \times S^2$, in which the AdS sector is written here by Poincar$\acute{e}$-type coordinates$(\hat{t}, \hat{r})$. It appears that the NHEG of the ultra-spinning black hole (\ref{UsMetric4d}) gives the well-known result which contains an AdS$_2$ sector. 
  The metric (\ref{NH4d}) similar to the horizon geometry, seems to be singular in $\theta=0, \pi$. We showed in previous subsections that these points are not truly singularities but they are some kinds of boundaries. Therefore the S$^2$ sector of NHEG (\ref{NH4d}) inherits the noncompactness characteristic and then it topologically would be a sphere with two punctures. 
 
 Metric (\ref{NH4d}) can be cast into the general form of the NHEG constructed in \cite{ChowCvetic-CFT:2008} which is calculated for the most general gauged extremal and stationary supergravity black holes as follows
 \begin{eqnarray}\label{NHGeneralform}
 ds^2&=&\Gamma(\theta)\bigg[-\rho^2 dt^2 + \frac{d\rho^2}{\rho^2} + \alpha(\theta) d\theta^2\bigg]+\gamma(\theta)(d\phi+k  \rho dt)^2,\\ \nonumber
 A^I&=& f(\theta)(d\phi+k \rho dt)+e^p r dt.
 \end{eqnarray}
 Where $\Gamma(\theta), \gamma(\theta)$ and $\alpha(\theta)$ are those extracted from (\ref{NH4d}). Additionally, to explore the behaviour of the gauge fields in the near horizon limit (\ref{NHG-Coordinate4d}) we should carry out a new gauge transformation $A^I \rightarrow A^I + d\Lambda^I$ on parameter $\Lambda$ \cite{CompereKerrCFT}.
 \begin{equation}\label{NHG-gauge2}
 \Lambda=\frac{\Phi_e^{I \, ext}}{\lambda}r_0 \hat{t},
 \end{equation}
 where $\Phi_e^{I \, ext}$ are the electrostatic potentials on the horizon in the extremal limit. This gauge transformation can be realized as a simple embeddings of a $U(1)$ gauge field in a higher-dimensional auxiliary spacetime \cite{CompereKerrCFT}. 
 
 %%%%%%%%%%%%%%%%%%%%%%%%%%%%%%%% Kerr/CFT description of ultra-spinning black hole %%%%%%%%%%%%%%%%%%%%%%%%%%%%%%%%%%%%
 \subsection{A quick review on Kerr/CFT}
 Here, we test the conjecture Kerr/CFT correspondence and its extensions for our noncompactness horizon black hole. Kerr/CFT explains that there exists a duality between the near horizon states of a $4$d extremal kerr black hole and a certain $d=2$ chiral conformal field theory \cite{GuicaHartmanStrominger-KerrCFT}. 
 The general NHEG (\ref{NHGeneralform}) by imposing a set of consistent boundary conditions admits an enhanced $SL(2, R)_R \times U(1)^n_L$ isometry group \cite{GuicaHartmanStrominger-KerrCFT, Barnich}. Indeed it includes all the symmetries of AdS$_2$ plus translations in $\varphi^i$ coordinates. The symmetry generators are given by the following Killing vectors fields
 \begin{equation}\label{symGenSL2U1}
\xi_1=\partial_t, \quad  \xi_2=t\partial_t - r \partial_r, \quad  \xi_3=\frac{1}{2}\big(\frac{1}{r^2}+t^2\big)\partial_t - t r \partial_r -\sum_{i=1}^{n}\frac{k_i}{r}\partial_\phi^i, \quad \quad  \bar{\xi}_i=\partial_\phi^i, 
 \end{equation}
One can write a relation between the $SL(2, R)$ and $U(1)^n$ symmetry generators \cite{Sheikh-JabbariHajianSeraj:2013}
\begin{equation}\label{SU2U1}
n^i\xi_i=k^i\bar{\xi}_i,
\end{equation}
where $n_i$ is the unit vector normal to AdS$_2$.

Therefore, because of the exact similarity between NHEG of our solutions (\ref{NHG-Coordinate4d}) and (\ref{NHGeneralform}), one can deduce that the metric (\ref{NH4d}) is invariant under diffeomorphisms generated by $\bar{\xi}_1$ and $\xi_{1,2,3}$. Therefore an asymptotic symmetry group $(ASG)$ associated to every consistent set of boundary conditions can be found. Asymptotic symmetries of (\ref{NH4d}) may contain diffeomorphisms $\zeta$ and a $U(1)$ gauge transformation such that \cite{HartmanStrominger:2009CFTDual}
 \begin{eqnarray}\label{ASG1}
 \delta_\zeta g_{\mu\nu}&=& \mathcal{L}_\zeta g_{\mu\nu},  \quad\quad    \delta_\zeta A_\mu= \mathcal{L}_\zeta A_\mu, \quad\quad  \delta_\Lambda A=d\Lambda.
 \end{eqnarray}
 The infinitesimal field variations are defined by $a_\mu=\delta A_\mu$ and $h_{\mu\nu}=\delta g_{\mu\nu}$. The combined transformation $(\zeta,\Lambda)$ has an associated charge $Q_{\zeta,\Lambda}$ defined in \cite{Barnich}, which generates the symmetry $(\zeta,\Lambda)$ under Dirac brackets. 
 
 The charge $Q_{\zeta,\Lambda}$ must be finite for all field variations, required to satisfy a consistent boundary condition. Now we choose the same boundary conditions as in \cite{GuicaHartmanStrominger-KerrCFT}. The method used in \cite{GuicaHartmanStrominger-KerrCFT} assumed that $\partial_\phi$ proportional to the zero mode of a nontrivial Virasoro algebra, as well as indicating the boundary conditions in terms of power law falloff of the components of the metric fluctuations $h_{\mu\nu}$ as
 \begin{eqnarray}\label{ASG-Boundary}
 \begin{pmatrix}
 h_{tt}=\mathcal{O}(r^2)& h_{t\varphi}=\mathcal{O}(1)& h_{t\theta}=\mathcal{O}(1/r)&h_{tr}=\mathcal{O}(1/r^2)\\
 
 & h_{\varphi\varphi}=\mathcal{O}(1)& h_{\varphi\theta}=\mathcal{O}(1/r)&h_{\varphi r}=\mathcal{O}(1/r)\\
 
 & & h_{\theta\theta}=\mathcal{O}(1/r)&h_{\theta r}=\mathcal{O}(1/r^2)\\
 
 & & &h_{r r}=\mathcal{O}(1/r^3)\\
 \end{pmatrix}.
 \end{eqnarray}
 Also, the boundary condition for the gauge field reads
 \begin{equation}\label{GaugeBoundary}
 a_\mu=\mathcal{O}(r,\frac{1}{r},1,1/r^2).
 \end{equation}  
 The most general diffeomorphisms $\zeta_m$ that preserve the above boundary conditions are given by
 \begin{eqnarray}\label{GeneralDiffeo}
 \zeta_\epsilon=\epsilon(\phi) \partial_\psi - r \epsilon^\prime(\phi) \partial_r, \quad \quad \quad \bar{\zeta}=\partial_t,
 \end{eqnarray}
 and the Virasoro algebra can be extracted as
 \begin{equation}\label{Virasoro1}
 i[\zeta_m,\zeta_n]=(m-n)\zeta_{m+n}, \quad \quad \epsilon_n(\phi)=-e^{-in \phi}.
 \end{equation}  
 The gauge field transformation $A$ under $\zeta_\epsilon$ does not satisfy the boundary condition (\ref{GaugeBoundary}). So to restore $\delta A_\varphi=\mathcal{O}(1/r)$ we must impose a suitable compensating $U(1)$ gauge transformation as $\Lambda=-f(\theta)\epsilon(\phi)$ \cite{HartmanStrominger:2009CFTDual}. The Virasoro algebra with vanishing central charge of ASG reads
 \begin{eqnarray}\label{Virasoro3}\nonumber
 [\Lambda_m,\Lambda_n]_\zeta&=&\zeta_m^\mu\partial_\mu\Lambda_n-\zeta_n^\mu\partial_\mu\Lambda_m,\\ 
 i[(\zeta_m,\Lambda_m),(\zeta_n,\Lambda_n)]_\zeta&=&(m-n)(\zeta_{m+n},\Lambda_{m+n}).
 \end{eqnarray}
 The gauge transformation $(\zeta_n,\Lambda_n)$ and charges $Q_n$ associated to them satisfy a similar algebra up to a central extension. Using (\ref{NHGeneralform}), one can derive the Dirac brackets algebra of $Q_n$ as \cite{Barnich,HartmanStrominger:2009CFTDual} 
 \begin{eqnarray}\label{ASG3} \nonumber
 i \big\{Q_{\zeta_\epsilon,\Lambda},Q_{\zeta_{\bar{\epsilon}},\tilde{\Lambda}}\big\}_{DB}&=&iQ_{(\zeta_\epsilon,\Lambda),(\zeta_{\bar{\epsilon}},\tilde{\Lambda})}-\frac{i k}{16 \pi}\int{d\theta d\varphi} \sqrt{\frac{\alpha(\theta)\gamma(\theta)}{\Gamma(\theta)}}\bigg(f(\theta) \Lambda \tilde{\epsilon}^\prime+\Gamma(\theta)\epsilon^\prime\tilde{\epsilon}^{\prime\prime}\\
 &+&[f(\theta)^2+\gamma(\theta)]\epsilon\tilde{\epsilon}^\prime-(\epsilon,\Lambda \leftrightarrow \tilde{\epsilon},\tilde{\Lambda})\bigg).
 \end{eqnarray}
 The algebra of the charge $Q_n$ associated to ASG generators $(\zeta_n,\Lambda_n)$ from (\ref{ASG3}) is
 \begin{equation}\label{Virasoro4}
 i\big\{Q_m,Q_n\big\}_{DB}=(m-n)Q_{m+n} + \frac{c}{12}(m^3-\alpha m) \delta_{m+n,0}.
 \end{equation}
We note that, $\alpha$ is a constant and can be absorbed to $Q_0$. The central charge $c_L$ has combinations of $k^{grav}$ and $k^{gauge}$ as $c=c_{grav}+c_{gauge}$. They can be nicely calculated in the manner described in \cite{HartmanStrominger:2009CFTDual,ChowCvetic-CFT:2008} as 
 \begin{equation}\label{CentarlCharge}
 c_{grav}=\frac{3k_i}{2\pi}\int_0^{\pi}d\theta \sqrt{\Gamma(\theta)\alpha(\theta)\gamma(\theta)}, \quad \quad c_{gauge}=0,
 \end{equation}
 where the constants $k_i$'s are given by corresponding Frolov-Thorne temperatures as below
 \begin{equation}\label{Frolov1}
 k_i=\frac{1}{2\pi T_i}.
 \end{equation}
 Here, $T_i$'s referred to the left- and right-moving temperatures of the quantum field theory coming from the Frolov-Thorne vacuum, are restricted to an extreme Kerr black hole \cite{FrolovThorne,GuicaHartmanStrominger-KerrCFT}. In order to determine these temperatures, one can use a scalar field expansion in terms of eigenstates of the asymptotic energy $E$ and angular momentum $J$
 \begin{equation}\label{FrolovVacume}
 \Phi=\sum_{E,J,l}^{} \phi_{E J l} \, e^{-i E \hat{t} + i J \hat{\varphi}} f_l(r,\theta).
 \end{equation}
 In the near horizon region (\ref{NHG-Coordinate4d}) we have
 \begin{equation}\label{FrolovVacumeNH}
 e^{-i E \hat{t} + i J \hat{\varphi}} = e^{-i (E-\Omega^0_H J)\hat{t} r_0/\lambda + i J \hat{\varphi}} =e^{-i n_R \hat{t} + n_L \hat{\varphi}},
 \end{equation}
 where $n_R= (E-\Omega^0_H J) r_0/\lambda$ and $n_L=J$ are the left and right charges associated to $\partial_\phi$ and $\partial_t$ in the near horizon metric. One can find a diagonal density matrix associated to the vacuum by tracing (\ref{FrolovVacume}) over the region inside the horizon. The Boltzmann weighting factor in the energy-angular momentum eigenbasis reads as 
  \begin{equation}\label{BoltzmanWeighting}
  e^{-\frac{(E-\Omega_H J)}{T_H}}
  \end{equation}
 In the nonrotating $(\Omega_H=0)$ case, this vacuum reduces to the Hartle-Hawking vacuum. One can hence write the Boltzmann factor in terms of  $n_L$, $n_R$ and $T_i$ as 
 \begin{equation}\label{FrolovVacumeBoltzman}
 e^{- \frac{(E-\Omega_H J)}{T_H}}=e^{-\frac{n_L}{T_L}-\frac{n_R}{T_R}}.
 \end{equation}
Dimensionless Frolov-Thorne temperatures $T_i$ where defined firstly for higher-dimensional Kerr-AdS black holes in \cite{LuMeiPope-KerrCFT} and are
 \begin{equation}\label{FrolovTRTL}
 T_L=\lim\limits_{r_+\rightarrow r_0}\frac{T_H^0}{\Omega^0_H-\Omega_H}=-\frac{\partial T_H / \partial r_+}{\partial \Omega_H / \partial r_+}\vert_{r_+=r_0}, \quad \quad \quad T_R=\frac{r_0}{\lambda}T_H\vert_{r_+=r_0}.
 \end{equation}
For an extremal solution, The right-temperature vanishes, while there are $[(d-1)/2]$ left-temperatures associated to the $2$d CFTs for each azimuthal $\varphi$ coordinate. However, the extreme Kerr black hole has vanishing Hawking temperature, but quantum fields states outside the horizon live in a thermal state.

We recall that (\ref{FrolovVacume}) can also be extended using the first thermodynamics law for a charged rotating black hole 
 \begin{equation}\label{FirstLaw} 
 TdS=dM-(\Omega_H dJ + \Phi dQ).
 \end{equation}
 Now by imposing the extremality constraint $T^{ex} dS=0$, it gets finally
 \begin{equation}
dS=\frac{d J}{T_L}+\frac{dQ}{T_e}.
 \end{equation}
 Also the Boltzmann factor can be extended to  
 \begin{equation}\label{BoltzmanGeneral} 
 e^{-n_R/T_R-n_L/T_L-Q/T_e}.
 \end{equation}
 %%%%%%%%%%%%%%%%%%%%%%%%%%%%%%%%%%%%%%%%%%%%%%%%%%%%%%%%%%%%%%%%%%%%%%%%%%%%%%%%%%%%%%%%%%%%%%%%%%%%%%%%%%%%%%%%%%%%%
 %%%%%%%%%%%%%%%%%%%%%%%%%%%%%%% UltraSpinning-CFT Correspondence  %%%%%%%%%%%%%%%%%%%%%%%%%%%%%%%%%%%%%%%%%%
 \subsection{Ultra-spinning/CFT description}
 In this section we shall study Kerr/CFT correspondence for our obtained extremal ultra-spinning $U(1)^4$ gauged supergravity black hole (\ref{UsMetric4d}). Our strategy besides confirming the existence of the CFT dual, will be to verify that the ultra-spinning limit and the Kerr-CFT limit commute with each other. It has been shown in \cite{HennigarKubiznakMann:2015} that the ultra-spinning limit commutes with the near horizon limit for $5$d Kerr-AdS and minimal gauged supergravity black holes. 
 
 Here, we consider two different procedures to explore Kerr/CFT for our ultra-spinning black hole. 
 i) Beginning with an extremal ultra-spinning black hole and then taking near horizon limit into account. 
 ii) Finding the near horizon limit of an extremal black hole and then applying the ultra-spinning limit afterwards. These two different ways are clearly shown in Fig 2.
 %%%%%%%%%%%%%%%%%%%%%%%%%%%%%%%%%%%%%%%%%%%%%%%%%%%%%%%%%%%%%%
 \begin{figure}[htp]
 	\centering
 	\begin{tabular}{ccc}
 		\includegraphics[width=.6\textwidth,height=.21\textheight]{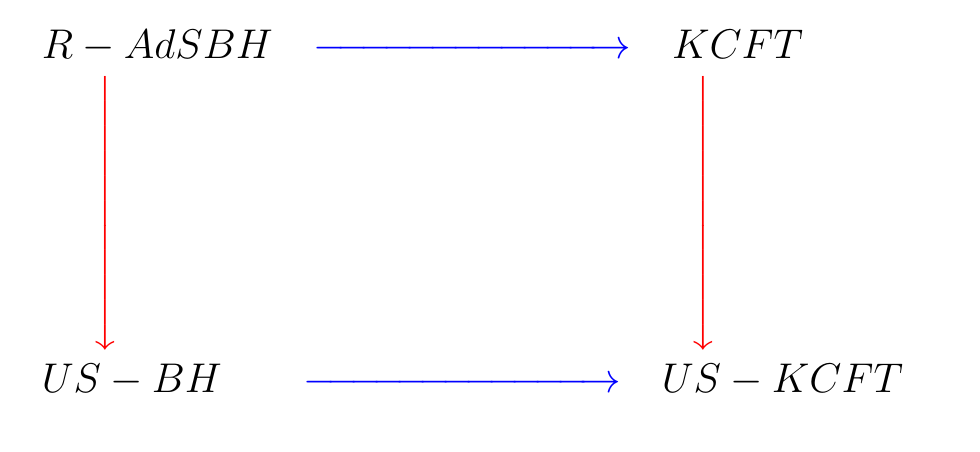}
 	\end{tabular}
 	\captionsetup{labelformat=empty}
 	\caption{{\bf Fig. 2.} This diagram illustrates two different order limits for a general rotating AdS black hole(R-AdS BH). Horizontal arrows (blue) represent the near horizon (NH) limit, to provide Kerr-CFT limit. Also the vertical ones (red) show the ultra-spinning (US) limit. We show that in both paths the resulting limit (US-KCFT) are exactly the same.  }
 \end{figure}\label{Fig2}
 %%%%%%%%%%%%%%%%%%%%%%%%%%%%%%%%%%%%%%%%%%%%%%%%%%%%%%%%%%%%%% 
 
 Let us here, start the lower path in Fig. 2. The resulting geometry is already obtained in (\ref{NH4d}). To find the central charge associated to metric (\ref{NH4d}), we need to write the extended version of the first law as 
 \begin{equation}\label{FirstLawUS4d}
 TdS=dM-\Omega_H dJ - \sum_{i}^{}{\Phi_i dQ_i} - K d\mu.
 \end{equation}
 Here, we consider $\mu$ as a thermodynamic parameter shown as a chemical potential \cite{HennigarKubiznakMann:2014}. It was explained in \cite{Herzog} that the compactified null length can be signified as a chemical potential. Since the new coordinate $\varphi$ is a compact null coordinate on the conformal boundary that becomes compactified by $\mu$, then $\mu$ is assumed to be a chemical potential, and its thermodynamic conjugate is denoted by $K$. Note that $\mu$ is a dimensionless quantity, so in Smarr formula there is no $K\mu$. Considering the extremality constraint on (\ref{FirstLawUS4d}) gives
 \begin{equation}\label{FirstLawUS4d2}
 TdS=-\big[ (\Omega_H-\Omega_H^{ex}) dJ + \sum_{i=1}^{4}{(\Phi_i-\Phi_i^{ex}) dQ_i} + (K-K^{ex}) d\mu \big].
 \end{equation}

 Then the Boltzmann factor (\ref{FrolovVacumeBoltzman}) in the extremal limit takes the following form
 \begin{equation}\label{BoltzmanUS4d2}
 e^{-n_R/T_R-n_L/T_L-\sum_{i=1}^{4} Q_i/T_{i,e}-\mu/T_\mu},
 \end{equation}
 where $n_R=(E-\Omega_H^{ex} J - \sum_{i}^{}\Phi_i^{ex} Q_i -K^{ex}\mu) r_0/ \lambda$ and $n_L=J$. Also $T_R$ and $T_L$ are given by (\ref{FrolovTRTL}). The Frolov-Thorne temperatures $T_e$ and $T\mu$ are defined as
 \begin{equation}\label{FroloVUS4d}
 T_{e,i}=- \frac{\partial T_H/ \partial r_+ }{\partial \Phi_i/ \partial r_+}|_{r_+=r_0},\quad\quad  T_{\mu}=-\frac{\partial T_H/ \partial r_+ }{\partial K/ \partial r_+}|_{r_+=r_0}.
 \end{equation}
 Now, using (\ref{FrolovTRTL}) and (\ref{FroloVUS4d}) we obtain following left- and right-moving temperatures
 \begin{eqnarray}\label{TL4d}
 T_R&=&0,\quad \quad \quad T_L=\frac{1}{2\pi k}=\frac{X\big[(r_0+q_1)(r_0+q_2)+\ell^2\big]}{2 \pi \ell (2r_0+q_1+q_2)}, 
 \end{eqnarray}
 and finally using (\ref{CentarlCharge}) we find central charge as 
 \begin{equation}\label{cUS4d}
 c=3\frac{\mu}{\pi} \frac{\ell (2r_0+q_1+q_2)}{X}.
 \end{equation} 
 
 We emphasis that the left-moving part of CFT identifies the quantum field states on the near horizon region (\ref{NH4d}). Namely, in the extremal limit the vacuum state in the bulk reduces to a mixed density matrix on the CFT side. The Boltzmann weighting factor reads
 \begin{equation}\label{density matrixUS4d}
 \rho=e^{-\frac{J}{T_L}-\sum_{i}^{}{\frac{Q}{T_{e,i}}}-\frac{\mu}{T_\mu}}.
 \end{equation}
 
 Indeed, the CFT dual of the generalised Hartel-Hawking vacuum has temperature $T_L$. Now we apply the thermodynamic Cardy formula that counts microstates of an unitary and modular invariance CFT at large $T$, and gives the entropy of CFT relating to its temperature and central charge \cite{Cardy}. 
 \begin{equation}\label{Cardy}
 S=\frac{\pi^2}{3}c_L T_L.
 \end{equation}
 This relation confirms that $c_L$ can be viewed as a measure of degrees of freedom and it determines the asymptotic density of states.
 
 Now, using  (\ref{TL4d}) and (\ref{cUS4d}) we obtain the microscopic entropy of the dual CFT 
 \begin{equation}
 S_{CFT}=\frac{\mu}{2} \big[ (r_+ +q_1)(r_+ + q_2)+\ell^2 \big],
 \end{equation}
 which agrees precisely with the macroscopic Bekenstein-Hawking entropy (\ref{EntropyUS4d}). So this remarkable result confirms the Kerr/CFT conjecture again.
 
 %%%%%%%%%%%%%%%%%%%%%%%%%%%%%%%%%%%%%%%%%%%%%%%%%%%%%%%%%%%%%%%%%%%%%%%%%%%%%%%%%%%%%%%%%%%%%%%%%%%%%%%%%%%%%%%%%%
 Now, we turn to the upper path in Fig. 2. In the first step we have to find the near horizon geometry of the metric (\ref{Metric4d)}) in the extremal limit, and then take the ultra-spinning limit on the NHEG. In \cite{ChowCvetic-CFT:2008} the NHEG of the metric (\ref{Metric4d)}) has been constructed. Its ultra-spinning limit can easily be calculated by replacing new coordinate $\varphi=\frac{\phi}{\Xi}$ and then setting $a \rightarrow \ell$ and finally compactifying the new coordinate $\varphi$ with period $\mu$. One can easily check that the final result is exactly the same as (\ref{NH4d}). So we can conclude that their CFT dual and central charge are the same too. 
 
 Hence these two results coming from procedures i) and ii) confirm that the ultra-spinning and the near horizon limits commute with each other for a $4$D gauged supergravity black hole solution.
%%%%%%%%%%%%%%%%%%%%%%%%%%%%%%%%%%%%%%%%%%%%%%%%%%%%%%%%%%%%%%%%%%%%%%%%%%%%%%%%%%%%%%%%%%%%%%%%%%%%%%%%%%%%%%%%%%

Another issue that can also be discussed is that, in black hole (\ref{UsMetric4d}), the angular momentum $J$ is not independent of the mass $M$. As we see in Eq. (\ref{thermoUS4d}), they are related by the chirality-type condition $M = J/ l$. We follow the thermodynamic interpretation as that in \cite{Klemm:2014}; thus we can define $L_0$ and $\tilde{L}_0$ in terms of $M$ and $J$ as
 \begin{equation}\label{L0}
 L_0=\frac{(M+J/l)}{2}, \quad \quad \quad   \tilde{L}_0=\frac{(M-J/l)}{2}.
 \end{equation}
So in order to derive the Smarr formula in terms of $L_0$ and $\tilde{L}_0$, we should consider $M=M(L_0,  \tilde{L}_0, Q , \mu) $. Since in our case $\tilde{L}_0$ vanishes, the first law becomes 
 \begin{equation}\label{FirstLawL0}
 TdS=(1-\Omega)dL_+-\sum_i \Phi dQ - K d\mu,
 \end{equation}
 and the chirality condition $M=J/l$ states that the black hole microstates can be explained by chiral excitations of a $3$D conformal field theory.

 %%%%%%%%%%%%%%%%%%%%%%%%%%%%%%%%%%%%%%%%%%%%%%%%%%%%%%%%%%%%%%%%%%%%%%%%%%%%%%%%%%%%%%%%%%%%%%%%%%%%%%%%%%%%%%%%%%
 %%%%%%%%%%%%%%%%%%%%$5$-dimensional $U(1)^3$ gauged supergravity black holes%%%%%%%%%%%%%%%%%%%%%%%%%%%%%%%%%%%%%%
 %%%%%%%%%%%%%%%%%%%%%%%%%%%%%%%%%%%%%%%%%%%%%%%%%%%%%%%%%%%%%%%%%%%%%%%%%%%%%%%%%%%%%%%%%%%%%%%%%%%%%%%%%%%%%%%%%%
 \section{$5$-dimensional $U(1)^3$ gauged supergravity black holes}\label{section-5d}
 In this section we consider a charged rotating black hole which is constructed as a solutions of $SO(6)$ gauged five-dimensional supergarvity, whose relevant part of bosonic action is given by \cite{Gunaydin-5dSupergravity}
 \begin{equation}\label{Action5d}
 S_{5d}=\frac{1}{16\pi}\int d^5x \sqrt{-g} \bigg( R-\frac{1}{2}\partial\overrightarrow{\phi}^2-\frac{1}{4}\sum_{i=1}^{3}X_i^{-2}(F^i)^2+\frac{4}{\ell^2}\sum_{i=1}^{3}X_i^{-1}+\frac{1}{24}
 \epsilon_{ijk}\epsilon^{\mu\nu\rho\sigma\lambda}F^i_{\mu\nu}F^j_{\rho\sigma}A^k_\lambda,
 \bigg),
 \end{equation}
 where $\overrightarrow{\phi}=(\phi_1,\phi_2)$, and
 \begin{equation}\label{E3}
 X_1=e^{-\frac{1}{\sqrt{6}}\phi_1-\frac{1}{\sqrt{2}}\phi_2},  \quad\quad\quad\quad X_2=e^{-\frac{1}{\sqrt{6}}\phi_1+\frac{1}{\sqrt{2}}\phi_2},  \quad\quad\quad\quad X_3=e^{\frac{2}{\sqrt{6}}\phi_1}.
 \end{equation}
 A six parameter family of solutions including three electric charges, two angular momenta and mass, which is the most general asymptotically AdS$_{5}$ black hole solution to this theory was constructed in \cite{Wu-General5GaugedSupergravity}. Here, we consider the $U(1)^3$ Cartan subgroup of $SO(6)$ with three charge parameters $\delta_I$ that satisfy $\delta_1=\delta_2:=\delta$ and $\delta_3=0$, as well as two independent rotation parameters. These four-parameters family of solutions was derived in \cite{ChongCveticLuPope-5dGaugedSupergravity}. Their metric is given by
 \begin{eqnarray}\label{Metirc5d}\nonumber
 ds^2&=&H^{-\frac{4}{3}}\bigg[-\frac{\Delta}{\rho^2}\big(dt-a\sin^2\theta\frac{d\phi}{\Xi_a}-b\cos^2\theta\frac{d\psi}{\Xi_b}\big)^2\\
 &+&\frac{C}{\rho^2}\big(\frac{ab}{f_3}dt-\frac{b}{f_2}\sin^2\theta\frac{d\phi}{\Xi_a}-\frac{a}{f_1}\cos^2\theta\frac{d\psi}{\Xi_b}\big)^2 +\frac{Z\sin^2\theta}{\rho^2}\big(\frac{a}{f_3}dt-\frac{1}{f_2}\frac{d\phi}{\Xi_a}\big)^2\\\nonumber
 &+&\frac{W\cos^2\theta}{\rho^2}
 (\frac{b}{f_3}dt-\frac{1}{f_1}\frac{d\psi}{\Xi_b})^2\bigg] + H^{\frac{2}{3}}\big(\frac{\rho^2}{\Delta}dr^2+\frac{\rho^2}{\Delta_\theta}d\theta^2\big),
 \end{eqnarray}
 where
 \begin{eqnarray}\label{Metirc5d2}\nonumber
 H&=&1+q/\rho^2,\quad \quad \quad \quad \quad \rho^2=r^2+a^2\cos^2\theta+b^2\sin^2\theta,  \\ \nonumber
 f_1&=&a^2+r^2, \quad \quad \quad  f_2=b^2+r^2, \quad \quad \quad \quad f_3=f_1 f_2+qr^2, \\\nonumber
 \Delta&=&\frac{1}{r^2}(a^2+r^2)(b^2+r^2)-2m+(a^2+r^2+q)(b^2+r^2+q)/\ell^2,\\
 \Delta_\theta &=& 1-\frac{a^2}{\ell^2}\cos^2\theta-\frac{b^2}{\ell^2}\sin^2\theta, \quad \quad \quad C=f_1f_2(\Delta+2m-q^2/\rho^2),\\\nonumber
 Z&=&-b^2C+\frac{f_2f_3}{r^2}[f_3-\frac{r^2}{\ell^2}(a^2-b^2)(a^2+r^2+q)\cos^2\theta],\\\nonumber
 W&=&-a^2C+\frac{f_1f_3}{r^2}[f_3+\frac{r^2}{\ell^2}(a^2-b^2)(b^2+r^2+q)\sin^2\theta],\\\nonumber
 \Xi_a&=&1-\frac{a^2}{\ell^2},\quad \quad \quad \Xi_b=1-\frac{b^2}{\ell^2}.
 \end{eqnarray}
 The gauge and scalar fields are 
 \begin{eqnarray}\label{Metirc5d-gauge}\nonumber
 A^1&=&A^2=\frac{\sqrt{q^2+2mq}}{\rho^2}(dt-a\sin^2\theta\frac{d\phi}{\Xi_a}-\cos^2\theta\frac{d\psi}{\Xi_b}),\\
 A^3&=&\frac{q}{\rho^2}(b\sin^2\theta\frac{d\phi}{\Xi_a}+a\cos^2\theta\frac{d\psi}{\Xi_b}).\\ \nonumber
 X_1&=&X_2=H^{-\frac{1}{3}}, \qquad X_3=H^{\frac{2}{3}}
 \end{eqnarray}
 The metric (\ref{Metirc5d}) is written in an asymptotic rotating frame. One can use a sort of coordinates to be asymptotically static frame (ASF) by taking $\phi=\tilde{\phi}+\frac{a}{\ell^2}t$ and $\psi=\tilde{\psi}+\frac{b}{\ell^2}t.$ 
 
 The Hawking temperature, entropy and the electrostatic potentials on the horizon in the asymptotically static frame are 
 \begin{eqnarray}\label{TemEnt5d} \nonumber
 T_H&=&\frac{2r_+^6+r_+^4(\ell^2+a^2+b^2+2q)-a^2b^2\ell^2}{2\pi r_+\ell^2[(r_+^2+a^2)(r_+^2+b^2)+qr_+^2]}, \quad  \quad
 S_{BH}=\frac{\pi^2[(r_+^2+a^2)(r_+^2+b^2)+qr_+^2]}{2r_+\Xi_a\Xi_b}, \\ 
 \Phi_1&=&\Phi_2=\frac{\sqrt{q^2+2mq} \, r_+^2}{(a^2+r_+^2)(b^2+r_+^2)+qr_+^2}, \quad \quad \quad \qquad \qquad \Phi_3=\frac{a q b}{(a^2+r_+^2)(b^2+r_+^2)+qr_+^2}.
 \end{eqnarray}
The Killing vector field that generates the Killing horizon is $K=\partial_t+\Omega_a^S \partial_\phi+\Omega_b^S \partial_\psi$, where $\Omega_a^S$ and $\Omega_b^S$ are the angular velocities on the horizon in (ASF). In ARF, the angular velocities on the horizon are written as
 \begin{equation}\label{OmegaARF5d} 
 \Omega_a^R=\Omega_a^S-\frac{a}{\ell^2}=\frac{\Xi_a a (r_+^2+b^2)}{(a^2+r_+^2)(b^2+r_+^2)+ q r_+^2}, \quad \quad \quad
 \Omega_b^R=\Omega_b^S-\frac{b}{\ell^2}=\frac{\Xi_b b (r_+^2+a^2)}{(a^2+r_+^2)(b^2+r_+^2)+ q r_+^2}.
 \end{equation} 
 
 Conserved charges including two angular momenta, two electric charges and mass as mentioned in \cite{ChongCveticLuPope-5dGaugedSupergravity} are
 \begin{eqnarray}\label{Charges5d} \nonumber
 J_a&=&\frac{\pi a (2m+q \Xi_b)}{4 \Xi_b \Xi_a^2}, \quad \quad \quad \qquad \qquad J_b=\frac{\pi b (2m+q \Xi_a)}{4 \Xi_a \Xi_b^2}, \\ 
 Q_1&=&Q_2=\frac{\pi \sqrt{q^2+2mq}}{4 \Xi_a \Xi_b}, \qquad \qquad \qquad Q_3=-\frac{\pi a b q}{4 \ell^2 \Xi_a \Xi_b},\\ \nonumber
 E&=&\frac{\pi \big[2m(2\Xi_a+2\Xi_b-\Xi_a\Xi_b)+q(2\Xi_a^2+2\Xi_b^2+2\Xi_a\Xi_b-\Xi_a^2\Xi_b-\Xi_b^2\Xi_a)\big]}{8 \Xi_a^2 \Xi_b^2}.	
 \end{eqnarray}

 %%%%%%%%%%%%%%%%%%%%%%%%%%%%%%%%%%%%%%%%%%%%%   ultra-spinning limit   %%%%%%%%%%%%%%%%%%%%%%%%%%%%%%%%%%%%%%%%%%%%
 \subsection{Ultra-spinning limit}
 
 We are now ready to perform the ultra-spinning limit on the metric (\ref{Metirc5d}), by following the same steps as in the previous section. This black hole rotates in two different directions $\phi$ and $\psi$, corresponding to the rotation parameters $a$ and $b$. But we are only allowed to take the ultra-spinning limit for one azimuthal direction which is selected as $\phi$ for us. Since the metric is already written in ARF, we begin by introducing a new azimuthal coordinate $\varphi=\phi / \Xi_a$. Then upon taking the $a \rightarrow \ell$ limit while we keep the parameter $b$ fixed, we will have 
 \begin{equation}\label{Delta5d}
 \Delta_{\theta}=\Xi_b \, \sin^2\theta.
 \end{equation}
 Hence we obtain the following new black hole solution as
 \begin{eqnarray}\label{MetricUS5d}\nonumber
 ds^2&=&\tilde{H}^{\frac{2}{3}}\bigg[-\frac{\tilde{\Delta}}{\rho^2}(dt-\ell\sin^2\theta d\varphi-b\cos^2\theta\frac{d\psi}{\Xi_b})^2\\ \nonumber
 &+&\frac{\tilde{C}}{\tilde{\rho}^2}(\frac{\ell b}{\tilde{f_3}}dt-\frac{b}{f_2}\sin^2\theta d\varphi-\frac{\ell}{\tilde{f_1}}\cos^2\theta\frac{d\psi}{\Xi_b})^2 	+\frac{\tilde{Z}\sin^2\theta}{\tilde{\rho}^2}(\frac{\ell}{\tilde{f_3}}dt-\frac{1}{f_2} d\varphi)^2\\	
 &+&\frac{\tilde{W}\cos^2\theta}{\tilde{\rho}^2}
 (\frac{b}{\tilde{f_3}}dt-\frac{1}{\tilde{f_1}}\frac{d\psi}{\Xi_b})^2 \bigg] + \tilde{H}^{\frac{2}{3}}(\frac{\tilde{\rho}^2}{\tilde{\Delta}}dr^2+\frac{\tilde{\rho}^2}{\Xi_b \sin^2\theta}d\theta^2),
 \end{eqnarray}
 where the $\tilde{H}$ , $\tilde{Z}$ , $\tilde{W}, \tilde{\Delta}$ and $\tilde{\rho}$ are given by (\ref{Metirc5d2}) in which $a \rightarrow \ell$. Also the new coordinate $\varphi$ can be compactified by $\varphi \sim \varphi + \mu$. The gauge potentials and scalar fields in this limit become 
 \begin{eqnarray}\label{USMetirc5d-gauge}\nonumber
 A^1&=&A^2=\frac{\sqrt{q^2+2mq}}{\tilde{\rho}^2}(dt-\ell\sin^2\theta d\varphi-\cos^2\theta\frac{d\psi}{\Xi_b}),\\
 A^3&=&\frac{q}{\tilde{\rho}^2}(b\sin^2\theta d\varphi+\ell\cos^2\theta\frac{d\psi}{\Xi_b}),\\ \nonumber
 X_1&=&X_2=\tilde{H}^{-1/3}, \quad\quad\quad\quad X_3=\tilde{H}^{2/3}.
 \end{eqnarray}
Now, it is easy to check that the metric (\ref{MetricUS5d}) and fields (\ref{USMetirc5d-gauge}) satisfy the equation of motion, and we can call (\ref{MetricUS5d}) as a new exact gauged supergravity black hole solution.

It is worth mentioning that taking the ultra-spinning limit in the $\psi$ direction (instead of $\phi$) is also carried out in the same way. But it is impossible to take the ultra-spinning limit in both directions simultaneously, because the $g_{\theta \theta}$ component in the metric (\ref{MetricUS5d}) will diverge in the $b \rightarrow \ell$ limit, and $1/\Xi_b$ divergence may not be absorbed into a new coordinate.
 
 %%%%%%%%%%%%%%%%%%%%%%%%%%%%%%%%%%%%%%%%%%%%%%%%%%%%%%%%%%%%%%%%%%%%%%%%%%%%%%%%%%%%%%%%%%%%%%%%%%%%%%%%%%%%%%%%%%
 \textbf{Horizon geometry: } Here, we find the induced metric on the horizon
 \begin{eqnarray}\label{ds2h-5d}
 ds^2_h&=&\tilde{H}^{-\frac{4}{3}}\bigg[\frac{\tilde{C}}{\tilde{\rho}^2}(\frac{b}{f_2}\sin^2\theta d\varphi+\frac{\ell}{\tilde{f_1}}\cos^2\theta\frac{d\psi}{\Xi_b})^2 +
 \frac{\tilde{Z}\sin^2\theta}{\tilde{\rho}^2}\frac{1}{f_2^2} d\varphi^2\\ \nonumber
 &+&\frac{\tilde{W}\cos^2\theta}{\tilde{\rho}^2}
 (\frac{1}{\tilde{f_1}}\frac{d\psi}{\Xi_b})^2 \bigg] +\tilde{H}^{\frac{2}{3}}\frac{\tilde{\rho}^2}{\Xi_b \sin^2\theta}d\theta^2.
 \end{eqnarray}
 
 As in the four-dimensional case, this metric seems to be ill defined in $\theta=0$ $( 0 \leq \theta \leq \pi/2)$. But to show that there is not problem near this point, one can perform the change of coordinates $k=l(1-\cos \theta)$, for small $k$, by which the horizon metric reads,
 \begin{equation}\label{eq12}
 ds^2_h=\frac{\rho_+}{\Xi_b}\left[\frac{dk^2}{4k^2}+4k^2\frac{(\ell^2+r_+^2)^3b^2\Xi_b}{r_+^2 \ell^2\rho_+^4}d\varphi^2+\frac{4bkm}{\rho_+ ^4}d\varphi d\psi \right] + \frac{2m}{\rho_+^2 \Xi_b} d \psi^2.
 \end{equation}
 For $\psi=$constant slices, it reduces to a metric of constant negative curvature on a quotient of the hyperbolic space  $\mathbb{H}^2$, indicating that the horizon is non-compact. In Fig. 3 the embedding topology for constant $\psi$ slices of the horizon are displayed for $\mu=2\pi$.
 
  \begin{figure*}
 	\centering
 	\begin{tabular}{ccc}
 		\includegraphics[width=0.3\textwidth,height=0.3\textheight]{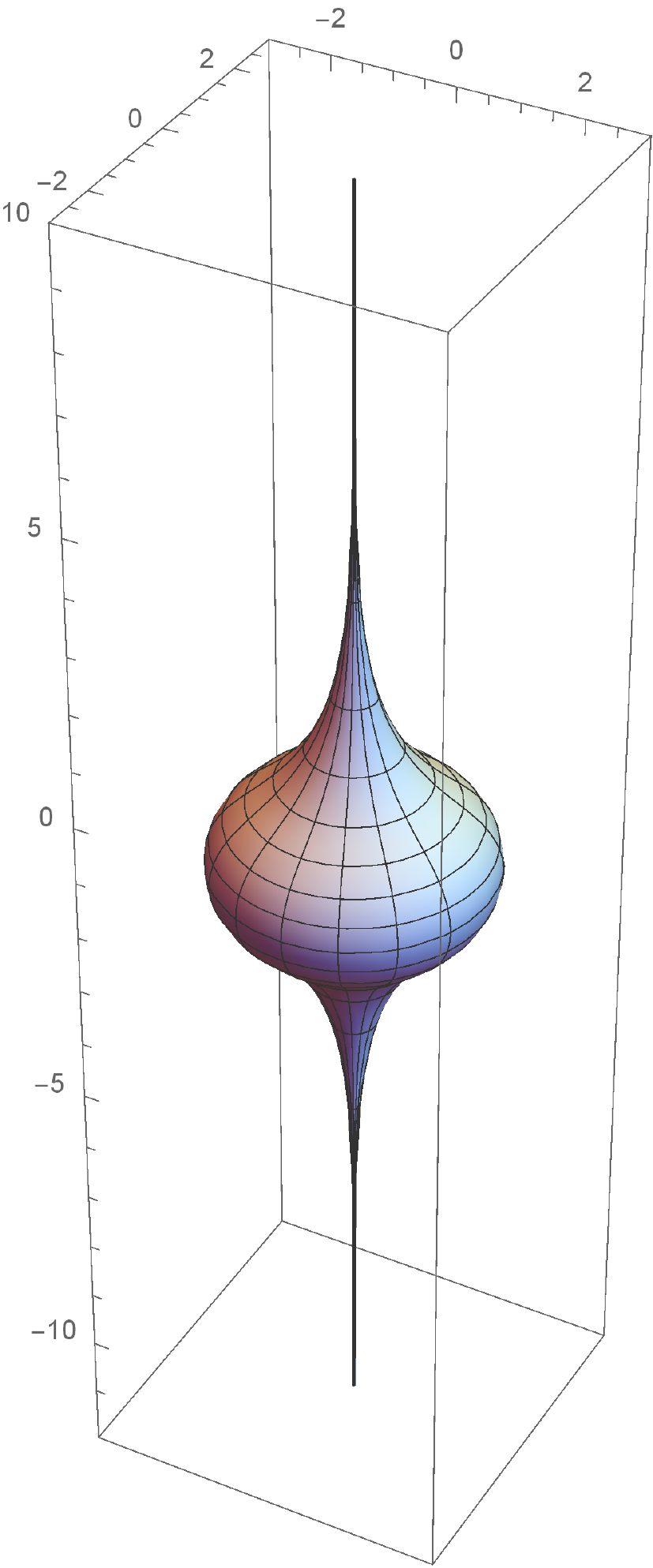} &
 		\includegraphics[width=0.3\textwidth,height=0.3\textheight]{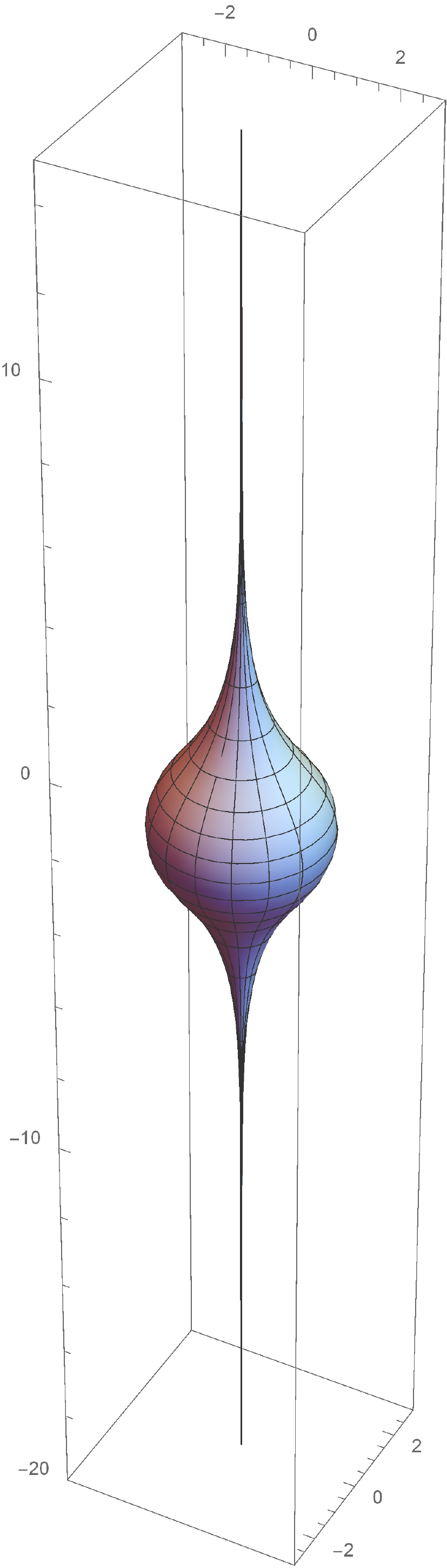}&
 		\includegraphics[width=0.3\textwidth,height=0.3\textheight]{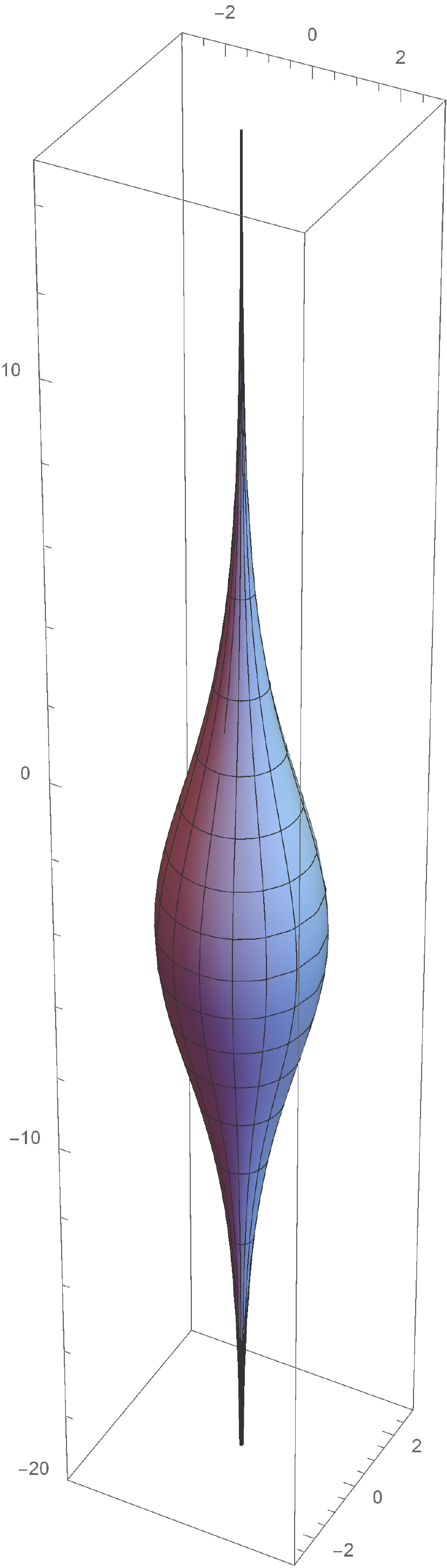}\\
 	\end{tabular}
 	\captionsetup{labelformat=empty}
 	\caption{{\bf Fig. 3. \bf Horizon embeddings in 5d}. Plots show the 2-dimensional $\psi=const.$ slices horizon (\ref{ds2h-5d}) embedded in $\mathbb{R}$$^3$ as a surface of revolution. we set $\mu = 2\pi, \, r_+ = 0.8, \, q=10$  and $\ell=1$ for all diagrams, and  $b=0.0$ (left), $b=0.5$ (middle), as well as $b=0.9$ (right). All cases show topology of a sphere with two punctures }\label{HorizonEmbedding5d}
 \end{figure*}
 %%%%%%%%%%%%%%%%%%%%%%%%%%%%%%%%%%%%%%%%%%%%%%%%%%%%%%%%%%%%%%%%%%%%%%%%%%%%%%%%%%%%%%%%%%%%%%%%%%%%%%%%%%%%%%%%%%
 \textbf{Conformal Boundary : } To gain a deeper understanding of the obtained geometry (\ref{MetricUS5d}), let us here take a look at its conformal boundary, with the conformal factor $\ell^2/r^2$. 
 \begin{eqnarray}\label{eq13}
 ds^2_{bdry}&=&-dt^2+\ell\sin^2\theta dt d\varphi+\frac{b\cos^2\theta}{\Xi_b}dt d\psi+\frac{\ell^2\cos^2\theta}{\Xi_b}d\psi^2\\ \nonumber
 &+&\frac{b\ell\sin^2\theta\cos^2\theta}{\Xi_b}d\varphi d\psi +\frac{\ell^2}{\Xi_b\sin^2\theta}d\theta.
 \end{eqnarray}
 The coordinate $\varphi$ is a null coordinate on the conformal boundary. Moreover, as before we analyze this metric near the pole $\theta=0$. Then for small $k$ we have
 \begin{equation}\label{eq14}
 ds^2_{bdry}=-dt^2+\frac{1}{\Xi_b}\big[\frac{dk^2}{4k^2} + 2k  dt d \varphi+ b dt d\psi + \ell^2 d \psi^2 + 2b k d\varphi d\psi \big]
 \end{equation}
 which for $\psi=$constant slices represents an AdS$_3$ written Hopf-like fibration over $\mathbb{H}^2$. It means that the pole $\theta=0$ is removed from the boundary and is indeed not part of the space-time indeed.

 %%%%%%%%%%%%%%%%%%%%%%%%%%%%%%%%%%%%%%%%%%%%%%%%%%%%%%%%%%%%%%%%%%%%%%%%%%%%%%%%%%%%%%%%%%%%%%%%%%%%%%%%%%%%%%%%%%
 \textbf{Thermodynamic quantities:} For new obtained ultra-spinning black hole (\ref{MetricUS5d}), we precisely derived the thermodynamic quantities including temperature, entropy and angular velocities on the horizon as well as electrostatic potentials on the horizon as
 \begin{eqnarray}\label{ThermoUS5d} \nonumber
 T_H&=&\frac{2r_+^6+r_+^4(2\ell^2+b^2+2q)-b^2l^4}{2\pi r_+\ell^2[(r_+^2+\ell^2)(r_+^2+b^2)+qr_+^2]}, \quad \quad \quad
 S_{BH}=\frac{\mu\pi[(r_+^2+\ell^2)(r_+^2+b^2)+qr_+^2]}{4r_+\Xi_b}, \\ \nonumber
  \Omega_a&=&\frac{\ell(b^2+r_+^2)}{(r_+^2+\ell^2)(r_+^2+b^2)+qr_+^2}r,\qquad \qquad \qquad      \Omega_b=\frac{b(r_+^4+2r_+^2\ell^2+r_+^2q+\ell^4)}{\ell^2(r_+^2+\ell^2)(r_+^2+b^2)+q\ell^2r_+^2},\\ 
 \Phi_1&=&\Phi_2=\frac{r_+^2\sqrt{q^2+2mq}}{(\ell^2+r_+^2)(b^2+r_+^2)+qr_+^2},   \quad \quad \quad  \quad   \Phi_3=\frac{q \ell b}{(\ell^2+r_+^2)(b^2+r_+^2)+qr_+^2}.
 \end{eqnarray}
 
 %%%%%%%%%%%%%%%%%%%%%%%%%%%%%%%%%%%%%%%%%%%%%%%%%%%%%%%%%%%%%%%%%%%%%%%%%%%%%%%%%%%%%%%%%%%%%%%%%%%%%%%%%%%%%%%%%%
 \textbf{Extremality under ultra-spinning limit: } 
 To ensure that the extremality condition is preserved under the ultra-spinning limit for this $5$d supergravity black hole, we similarly explore two different ways as mentioned in Fig. 2. So, we first obtain the extremality conditions of main metric (\ref{Metirc5d}) with a horizon at $r=r_0$ using Eq. (\ref{TemEnt5d}) by imposing $\Delta=0$ and $T_H|_{r=r_0}=0$. We find two following conditions
\begin{eqnarray}\label{ExCondition5d}
  q&=&- a b + \frac{r_0^2\sqrt{a^2+b^2+\ell^2+2r_0^2}}{\ell}, \\ \nonumber
  m&=&\frac{1}{2}\frac{\ell^4(a^2+r_0^2)(b^2+r_0^2)+r_0^2(Y- a^2 \ell)(Y - b^2 \ell)}{r_0^2 \ell^4},
\end{eqnarray}
where $Y=a b \ell - r_0^2\sqrt{a^2+b^2\ell^2+2r_0^2}$.

Now, we find the extremality conditions of our obtained ultra-spinning black hole using (\ref{ThermoUS5d})
\begin{eqnarray}\label{ExtUS5d}
 q&=&-b \ell + \frac{r_0^2\sqrt{b^2+2(\ell^2+r_0^2)}}{\ell} \\ \nonumber
 m&=&\frac{1}{2}\frac{\ell^4(b^2+r_0^2)(\ell^2+r_0^2)+r_0^2(Y_s-\ell^2(b-\ell))(Y_s + b \ell(b-\ell))}{r_0^2 \ell^4}
\end{eqnarray}
where $Y_s=r_0^2(\ell+\sqrt{b^2+2(\ell^2+r_0^2)})$. 
It is easy to check that Eq. (\ref{ExtUS5d}) will be achieved by taking $a \rightarrow l$ limit on (\ref{ExCondition5d}). Hence as the previous case, the extremality condition and the ultra-spinning limit commute with each other for five-dimensional $U(1)^3$ gauged supergavity class of black holes.
 
 %%%%%%%%%%%%%%%%%%%%%%%%%%%%%%%%%%%%%%%%%%%%%%%%%%%%%%%%%%%%%%%%%%%%%%%%%%%%%%%%%%%%%%%%%%%%%%%%%%%%%%%%%%%%%%%%%%
 \subsection{Near horizon geometry}
  For an extremal solution of an ultra-spinning black hole (\ref{MetricUS5d}) with a horizon at $r=r_0$, we have found the extremality conditions (\ref{ExtUS5d}). Then one can write the near horizon expansion as 
 \begin{equation}
 \Delta=\tilde{X}(r-r_0)^2+\mathcal{O} (r-r_0)^3,
 \end{equation}
 where
 \begin{equation}\label{V5d}
 \tilde{X}=2+\frac{3 \ell^2 b^2}{r_0^4}+\frac{1}{\ell^2}(6 r_0^2+b^2+2 q).
 \end{equation}
 To obtain the near horizon geometry, we use the coordinate changes similar to (\ref{NHG-Coordinate4d}) as
 \begin{equation}\label{NHG-Coordinate5d}
  r=r_0(1+\lambda \hat{r}), \quad \quad \quad \varphi=\hat{\varphi}+\Omega_a^0 \hat{t}, \quad \quad  \psi=\hat{\psi}+\Omega_b^0\hat{t}, \quad \quad  t=\frac{\hat{t}}{2\pi T^{\prime \, 0}_H r_0 \lambda}, \quad\quad \theta=\hat{\theta}.
 \end{equation}
 After applying the scaling parameter $\lambda \rightarrow 0$, the near horizon geometry in terms of the vielbeins takes the form of
 \begin{equation}\label{NHG-Metric5d}
 ds^2=H_0^{2/3}\frac{\rho_0^2}{\tilde{X}}\big(-\hat{r}^2 d\hat{t}^2 + \frac{d\hat{r}^2}{\hat{r}^2}\big)  + F(\theta)d\hat{\theta}^2 + \sum_{i=1}^{2}\hat{e}^i\hat{e}^i,
 \end{equation}
 where $F(\theta)=H_0^{2/3}\frac{\rho_0^2}{\sin^2\theta \, \Xi_b}$ and vielbeins are
 \begin{eqnarray}\label{Vielbeins1} \nonumber
 \hat{e}^1=\alpha_1 \hat{e}_1 + \alpha_2 \hat{e}_2, \quad \quad \hat{e}^2=\beta_1 \hat{e}_1 + \beta_2 \hat{e}_2, \\
 \hat{e}_1=d\hat{\varphi} + k_\varphi \hat{r} dt, \quad \quad \hat{e}_2=d\hat{\psi} + k_\psi \hat{r} dt,
 \end{eqnarray}
 and
 \begin{eqnarray}\label{Vielbeins2} \nonumber
 k_\varphi&=&\frac{2 \ell [(r_0^2+b^2)^2+qb^2]}{\tilde{X}[(r_0^2+\ell^2)(r_0^2+b^2)+qr_0^2] r_0}, \qquad \quad k_\psi=\frac{2b \, \Xi_b[(r_0^2+\ell^2)^2+q \ell^2]}{\tilde{X} [(r_0^2+\ell^2)(r_0^2+b^2)+qr_0^2]  r_0},\\ \nonumber
 \alpha_1&=&H_0^{-2/3}\frac{(r_0^2+\ell^2+q)\sin\theta }{\sqrt{\rho_0^2(1-\Xi_b \sin^2\theta)}}, \qquad \qquad \quad \alpha_2=-H_0^{-2/3}\frac{(r_0^2+b^2+q)\,\ell\,b\sin\theta }{\sqrt{\rho_0^2(1-\Xi_b \sin^2\theta)}}, \\\nonumber
 \beta_1&=&H_0^{-2/3}\frac{ b [(r_0^2+l^2)\rho_0^2+q r_0^2]\sin^2\theta }{r_0\,\rho_0^2 \sqrt{1-\Xi_b \sin^2\theta}}, \qquad  \quad  \beta_2=H_0^{-2/3}\frac{ \ell [(r_0^2+l^2)\rho_0^2+q r_0^2]\cos^2\theta }{r_0\,\rho_0^2 \, \Xi_b \sqrt{1-\Xi_b \sin^2\theta}}. \\ \nonumber
 \end{eqnarray}

 \subsection{Ultra-spinning/CFT description}
 In order to find the CFT dual of obtained $5$d ultra-spinning black hole (\ref{MetricUS5d}), as it was done through the previous section, we can follow two different paths of Fig. 2. We firstly start from the lower path. In (\ref{NHG-Metric5d}) we represented the NHEG of ultra-spinning solution. Also for the current case we choose the same boundary conditions as in \cite{GuicaHartmanStrominger-KerrCFT}, 
 \begin{eqnarray}\label{ASG-Boundary5d}
 \begin{pmatrix}
 \mathcal{O}(r^2)& \mathcal{O}(1)& \mathcal{O}(1)& \mathcal{O}(1/r)& \mathcal{O}(1/r^2)\\
 
 & \mathcal{O}(1) & \mathcal{O}(1) &\mathcal{O}(1/r) & \mathcal{O}(1/r)\\
 
 & & \mathcal{O}(1) &\mathcal{O}(1/r) & \mathcal{O}(1/r)\\
 
 & & & \mathcal{O}(1/r)& \mathcal{O}(1/r^2)\\
 & & & & \mathcal{O}(1/r^3)\\
 \end{pmatrix},
 \end{eqnarray}
 using the basis $(\hat{t},\hat{\varphi},\hat{\psi},\hat{\theta},\hat{r})$. One can then show that the five-dimensional near horizon geometry (\ref{NHG-Metric5d}) provides two copies of commuting Virasoro algebras \cite{LuMeiPope-KerrCFT}, generated by a pair of commuting diffeomorphisms
 \begin{eqnarray}\label{diffeo}
 \zeta_\varphi=-e^{in \varphi} \partial_\varphi- i n r e^{-in \varphi}  \partial_r, \qquad 
 \zeta_\psi=-e^{in \psi} \partial_\psi- i n r e^{-in \psi}  \partial_r.
 \end{eqnarray}
 These diffeomorphisms leading us to an asymptotic symmetry algebra of transformations that satisfy above boundary conditions. In \cite{ChowCvetic-CFT:2008} a general form for NHEG of rotating black holes in $d=2n+1$ dimensions was constructed 
 \begin{eqnarray}\label{NHGgeneral}
 ds^2&=&\Gamma(y)\big(-\rho^2 dt^2 + \frac{d\rho^2}{\rho^2}\big)+\sum_{\alpha=1}^{n-1}F_\alpha dy_\alpha^2 + \sum_{i,j=1}^{n-1}\tilde{g}_{ij} \tilde{e}_i  \tilde{e}_j, \\ \nonumber
 \tilde{e}_i&=&d\phi_i+k_i \rho dt, \quad \quad \quad k_i=\frac{1}{2 \pi T_i}. \quad \quad \quad
 \end{eqnarray}
 The NHEG of our new solution is exactly recast in this form. The metric (\ref{NHGgeneral}) has $(n-1)$ copies of the Virasoro algebra. The central charges in this form are given by \cite{ChowCvetic-CFT:2008}
 \begin{eqnarray}\label{Centralgeneral}
 c_i=\frac{3}{2\pi}k_i \int{d^{n-1}y_\alpha \bigg(det \tilde{g}_{ij} \prod_{\alpha=1}^{n-1} F_\alpha\bigg)^{1/2} } \int {d \phi_1 \dots d \phi_{n-1}}.
 \end{eqnarray}
 Then, for (\ref{NHG-Metric5d}) we obtain two central charges $c_1$ and $c_2$ associated to diffeomorphisms $\partial_{\hat{\varphi}}$ and $\partial_{\hat{\psi}}$ respectively
 \begin{eqnarray}\label{CentralgeneralUS5d}
 c_1&=&\frac{3 k_\varphi}{8 \pi} \int \sqrt{F(\theta) (\alpha_2\beta_1-\alpha_1\beta_2)} d\theta d\varphi d\psi=\frac{3 \mu \ell [(r_0^2+b^2)^2+qb^2]}{\tilde{X} \Xi_b r_0^2}, \\ 
 c_2&=&\frac{3 k_\psi}{8 \pi} \int \sqrt{F(\theta) (\alpha_2\beta_1-\alpha_1\beta_2)} d\theta d\varphi d\psi=\frac{3 \mu b [(r_0^2+\ell^2)^2+q \ell^2]}{\tilde{X}  r_0^2}. \nonumber
 \end{eqnarray}
 We note that the central charges contain the chemical potential $\mu$, which comes from compactification of the new azimuthal coordinate. Then similar to the $U(1)^4$ case we can write the first law of thermodynamics in its extremality limit, appearing as 
 \begin{equation}
 TdS=-[(\Omega_\varphi-\Omega_\varphi^{ex})dJ_\varphi+(\Omega_\psi-\Omega_\psi^{ex})dJ_\psi+\sum_{i=1}^{3} (\Phi_i-\Phi_i^{ex}dQ_i)+(K-K^{ex})d\mu],
 \end{equation}
 and the Boltzmann weighting factor for this case reads
 \begin{equation}
 e^{-n_R/T_R-n_\varphi/T_\varphi-n_\psi/T_\psi-\sum_{i=1}^{4} Q_i/T_{t,\, i}},
 \end{equation}
 where $n_R=(E-\Omega_\varphi^{ex}J_{Ex}-\Omega_\psi^{ex}J_{\psi}-\sum \Phi^{ex}_i Q_i)r_0/\lambda$,  $n_\varphi=J_\varphi$ and $n_\psi=J_\psi$. We extract again the Frolov-Thorne temperatures 
 \begin{eqnarray}
 T_R&\equiv& \frac{T_H r_0}{\lambda}=0,\\ \nonumber
 T\varphi&=&-\frac{\partial T_H/ \partial_{r_+}}{\partial \Omega_\varphi/\partial_{r_+}}|_{ex}=\frac{1}{2 k_\varphi}, \quad \quad \quad T_\psi=-\frac{\partial T_H/\partial_{r_+} }{\partial \Omega_\psi/\partial_{r_+}}|_{ex}=\frac{1}{2 k_\psi},  \nonumber
 \end{eqnarray}
 where $K_\varphi$ and $K_\psi$ are given by (\ref{Vielbeins2}). Ultimately the CFT entropy using the Cardy formula is computed as
 \begin{equation}
 S_{CFT}=\frac{\pi^2}{3}c_1 T_\varphi+\frac{\pi^2}{3}c_2 T_\psi=\frac{\mu\pi[(r_+^2+\ell^2)(r_+^2+b^2)+qr_+^2]}{4r_+\Xi_b},
 \end{equation}
 which is exactly the same as the Bekenstein-Hawking entropy (\ref{ThermoUS5d}).

 Now, we examine CFT duality via the upper path of Fig 2. The near horizon geometry of metric (\ref{Metirc5d}) in extremal limit was studied in \cite{ChowCvetic-CFT:2008}. One can easily check that by performing the ultra-spinning technique based on their result, our calculated NHEG (\ref{NHG-Metric5d}) will exactly be held. It means that we can again confirm that the near horizon limit commutes with the ultra-spinning limit for this $5$D black hole. Hence the CFT dual of $U(1)^3$ gauged supergravity black hole (\ref{Metirc5d}) gives us the same result for both the upper and lower paths of Fig. 2. 
 
 \section{Discussion}\label{Discussion}
 For better understanding of the physics of gauged supergravity black holes in large angular momentum, we have employed the novel ultra-spinning (super-entropic) limit proposed in  \cite{HennigarKubiznakMann:2014} to generate new exact supergravity black hole solutions. In particular, we use this simple ultra-spinning technique for four-dimensional singly spinning $U(1)^4$ and five-dimensional doubly spinning $U(1)^3$ gauged supergravity black holes. Our obtained black holes for both cases have an unusual horizon that is noncompact but with finite area. However in the first glance we see singularities in the coordinate angles $\theta=0,\pi$. We have shown that these poles are not parts of the spacetime and can be viewed as a sort of boundary that introduces punctures to the spacetime, providing a non-compact horizon. In \cite{HennigarKubiznakMann:2014,HennigarKubiznakMann:2015} it was shown that a relation between this kind of ultra-spinning limit and super-entropic black holes can be found by exploring the thermodynamics behavior of the obtained black holes in the context of the extended phase space thermodynamics. In our upcoming work we shall study the properties of these obtained ultra-spinning black holes in the extended phase space to find the range of parameter space, giving us super-entropic black holes. 
 
 Also we have shown that the extremality conditions and the near horizon limit are preserved under ultra-spinning limit, demonstrating that they commute with the ultra-spinning limit. We have also presented the NHEG of both ultra-spinning gauged black holes despite the noncompactness of their horizons, possessing an AdS$_2$ sector and a $S^{d-2}$ with two punctures. The appearance of the AdS$_2$ factor in the NHEG prompts us to explore whether these unusual new solutions exhibit the well-defined Kerr/CFT correspondence. We have also investigated kerr/CFT for both cases. Assuming the Cardy formula, we have shown that the microscopic entropy of the dual CFT for both new ultra-spinning gauged supergravity black holes in four and five dimensions precisely agrees with its Bekenstein-Hawking entropy. Recently a formula as a higher-dimensional generalization of the Cardy formula for the large energy limit of generic CFT has been presented in \cite{Shaghoulian}, which relates the entropy of a CFT to the vacuum energy on $S^1 \times \mathbb{R}^d$. Now, there is a curiosity to explore, whether this formula can reproduce the entropy of the ultra-spinning black holes at a high temperature.
 
 A further direction for our next research will be an investigation of applying a hyperboloid membrane limit \cite{CaldarelliEmparan:2008} to these supergravity black holes and their ultra-spinning versions to generate other new exact solutions.

 \section*{Acknowledgments}
 We would like to thank M. M. Sheikh-Jabbari for fruitful discussions, insights, and helpful comments on the draft. Also we like to thank D. Klemm and especially M. H. Vahidinia for their useful discussions. S. M. N. wishes to acknowledge Saeedeh Rostami for many comments, and the organizers of the “Recent Trends in String Theory and Related Topics” workshop, held in May 2016 in Tehran, at which some preliminary results of this work were presented. S. M. N. would also like to thank the members of the quantum gravity group at the Institute for Research in Fundamental Sciences (IPM) for useful discussions, as well as the IPM for hospitality while this project was completed.

%\vskip2mm

%

\end{document}